\documentclass[12pt]{article}
\usepackage[dvips]{color}
\def\theequation{\arabic{section}.\arabic{equation}}

\renewcommand{\title}[1]{\begin{center}\bf\Large #1\end{center}}

\renewcommand{\author}[1]{\begin{center}\large #1\end{center}}

\usepackage{latexsym,amsfonts}
\newcommand{\rr}{\mathbb{R}}
\newcommand{\nn}{\mbox{\bf \scriptsize c}}
\usepackage{cite,epsf}
\usepackage{graphicx,array}
\def\beq{\begin{equation}}
\def\eeq{\end{equation}}
\def\bea{\begin{eqnarray}}
\def\eea{\end{eqnarray}}

\begin{document}

\hspace{11.6cm}{\bf HU-EP-07/65}

\vspace{5mm}

\title{Operator Approach to Boundary Liouville Theory}

\vspace{5mm}

\author{
Harald Dorn ${}^a$
and George Jorjadze ${}^{a,\,b}$\\
{\small${}^a$Institut f\"ur Physik der
Humboldt-Universit\"at zu Berlin,}\\
{\small Newtonstra{\ss}e 15, D-12489 Berlin, Germany}\\
{\small${}^b$Razmadze Mathematical Institute,}\\
{\small M.Aleksidze 1, 0193, Tbilisi, Georgia}}

\vspace{5mm}

\begin{abstract}
\noindent 
We propose new methods for calculation of 
the discrete spectrum, the reflection amplitude and the
correlation functions of boundary Liouville theory on a strip 
with Lorentzian signature.
They are based on the structure of the vertex operator
$V=e^{-\varphi}$ in terms of
the asymptotic operators. The methods first are tested
for the particle dynamics in the Morse potential, 
where similar structures appear. 
Application of our methods to 
boundary Liouville theory reproduces the known results
obtained earlier in the bootstrap approach, 
but there can arise a certain extension
when the boundary parameters are near to critical values.
Namely, in this case we have found up to four different 
equidistant series of discrete spectra, 
and the reflection amplitude is modified respectively.
\end{abstract}

\vspace{15mm}

\vspace{5mm}

Hf; 11.25. Pm; 11.25. -w;

\newpage

\section{Introduction}

Boundary Liouville theory (BLT)  has been  studied intensively 
\cite{GN} in the beginning of  the  80's in parallel to the periodic 
\cite{BCT} and
unbounded \cite{D'H-J} cases. 
The main motivation for those investigations was Polyakov's 
non-critical string theory, 
but due to the rich integrable structure of Liouville theory,
the obtained results contributed to other areas 
(2d conformal field theory, 2d gravity, quantum groups, $etc$) as well.
The interest in  the boundary theory was renewed by the end of the 
90's in connection to the  dynamics of branes. 
The new results of the last decade 
\cite{FZZ, T, ZZ, M, BRZ} were obtained mostly by the bootstrap method in 
the Euclidean formulation of the theory, whereas the approaches 
of  the  80's mainly concerned  the space-time with Lorentzian
signature.  

In this paper we continue the investigation of BLT started in \cite{DJ}. 
We study the theory given on a strip with  Lorentzian signature
and use the operator approach in the Heisenberg picture. 
In \cite{DJ} we have constructed the vertex operator $V$, corresponding 
to the Liouville field exponential $e^{-\varphi}$. 
It is the simplest vertex operator, which is the basic building block  
in the vertex operators calculus. 
This operator can also be used in calculations
of the scattering data of the theory. The operator $V$ was constructed 
similarly to the periodic case \cite{BCT,OW} using a free-field
parameterization, the conformal symmetry and the conditions of 
locality and Hermiticity. The parameterizing field in \cite{DJ} was 
chosen to be the $in$-field of the theory.
The corresponding Liouville field configurations describe scattering
processes, and they belong to the hyperbolic monodromy.

For certain values of the boundary parameters the theory also
contains  bound states, which are  absent  in the periodic 
case. The monodromy for those field configurations switches to the
elliptic one and the free-field parameterization fails there. More
precisely, the free-field becomes complex with non-linearly related
real and imaginary parts. This case was studied in \cite{GN} using a
parameterization in terms of two fields, but the fields were not
completely free and an exact description of the Hilbert space, like
one has for the hyperbolic monodromy, is missing. 

The aim of the present paper is to explore the structure of the
vertex operator $V$ and  to use it for  the calculation 
of the reflection amplitude, the discrete spectrum and the correlations 
functions. These calculations can be compared with the results 
obtained by the bootstrap method in the Euclidean formulation 
of the theory \cite{FZZ, T, ZZ, M, BRZ}. The connection between 
the Euclidean and
Minkowskian formulations of Liouville theory are well established
in the periodic case \cite{DO, ZZ1, T1, JW} and it is a challenge to
find the relation for the boundary theory as well. Due to the rich spectral
picture of BLT, the operator approach seems to be appropriate and
becomes indeed effective.

The paper is organized as follows: In Section 2 we first briefly summarize
the main results of \cite{DJ}. Then we introduce a generic vertex operator
$V_\alpha$ given as an expansion in powers of the screening charges. We
calculate  the  first few coefficients of this expansion and 
establish the relation
between the parameters of the operator $V$ and the mass and boundary 
parameters of the theory.   
In Section 3 we consider particle dynamics in the Morse
potential as a zero mode approximation to BLT.
Here we develop a new scheme for quantum mechanical calculations.
Namely, we show how from the structure of the  operator $V$ one can read 
off the discrete spectrum of the theory. We also present a method for 
calculation
of the phase of the reflection amplitude providing its integral
representation. In addition, we calculate correlation functions. 
These methods are
applied to BLT in Section \nolinebreak 4. First we 
project the operator $V$ on the vacuum sector, where it becomes  
similar to the corresponding operator of the particle model
in the Morse potential.   
We find the discrete spectrum of the vacuum sector and 
calculate the reflection amplitude. 
Then we calculate the 1-point function corresponding to our vertex operator,
and   compare our calculations to the results obtained by the bootstrap 
method.
 In the last section we discuss open problems of the operator
approach, such as the quantization of the 
bound state sector, the S-matrix, $etc$.
Four appendices contain technical details  and useful formulae.

\setcounter{equation}{0}

\section{Free-field description of BLT}

\subsection{The classical theory}

The Liouville field $\varphi(\tau,\sigma)$ is given on the strip
$(\tau,\sigma)~$: $\tau\in\rr ,~\sigma\in(0,\pi)$ and its dynamics is
described by the action
\begin{equation}\label{S}
S=\frac{1}{2\pi}\int d\tau\int_0^\pi d\sigma
\left[(\partial_\tau\varphi)^2-(\partial_\sigma\varphi)^2-
4m^2e^{2\varphi}\right]-\frac{2m}{\pi}\,\int d\tau
\left[l\,e^{\varphi(\tau,0)}+ r\,e^{\varphi(\tau,\pi)}\right]~,
\end{equation}
where $l$ and $r$ are the boundary parameters in units of $m$
($m>0$). Note that the action used in \cite{DJ} differs from (\ref{S})
 by the factor $\pi$.

The variation of (\ref{S}) yields the Liouville equation and the
boundary conditions
\begin{equation}\label{FZZT-B}
\partial_\sigma\varphi|_{\sigma=0}= 2ml
\,e^\varphi|_{\sigma=0}~~~~~~~~~
\partial_\sigma\varphi|_{\sigma=\pi}=- 2mr
\,e^\varphi|_{\sigma=\pi}~.
\end{equation}
These conditions provide $\bar T(\tau)=T(\tau)$ and
$T(\tau+2\pi)=T(\tau)$, where
$T(x)=(\partial_x\varphi)^2-\partial^2_{xx}\varphi$ and $\bar
T(\bar x)=(\partial_{\bar x}\varphi)^2-\partial^2_{\bar x\bar
x}\varphi$ are the chiral ($x=\tau+\sigma$) and the anti-chiral
($\bar x=\tau-\sigma$) components of the energy-momentum tensor,
respectively. Due to the periodicity of $T(x)$ the description of
the boundary theory becomes similar to the periodic case. Namely,
one can classify the Liouville fields by the orbits of the group of
diffeomorphisms of $S^1$ \cite{BFP}.

The Liouville field exponential $V=e^{-\varphi}$ satisfies the
equation $\partial^2_{xx}V=T(x)V$ and for a constant energy-momentum
tensor $T(x)={T_0}$ it is given by
\begin{equation}\label{V0}
V_0=\frac{2m}{p\sinh \pi p}\, \left[\sqrt{\Lambda(l,r;p)}\,\cosh p(\tau-\tau_0)+
l\,\cosh p(\sigma-\pi) +r\,\cosh p\sigma\right]~,
\end{equation}
where $\tau_0$ is an arbitrary constant, $~p=2\sqrt {T_0}$ and
\begin{equation}\label{u}
 \Lambda(l,r;p)={l^2+r^2
+2l\,r\,\cosh\pi p +\sinh^2 \pi p }~.
\end{equation}
For $T_0>0$ the solutions (\ref{V0}) describe a scattering process
with $\varphi\rightarrow -\infty$ at the time asymptotics $\tau
\rightarrow \pm\infty$, whereas the case $T_0<0$ ($p=i\theta$)
corresponds to bound states oscillating in time.

Necessary conditions for positivity of $V_0$ in the whole bulk 
$\sigma\in(0,\pi)$ are: $l\geq  -1\,$ and $\,\,r\geq -1$. In addition,
there are restrictions on allowed values of $T_0$. Here one has to
distinguish two cases: $l+r\geq0$ and $l+r<0$. The first case 
describes only scattering processes: $T_0$ can not take 
negative values, but all positive values are allowed. 
Bound states come into the game only for $l+r<0$. In this case 
$T_0\geq -\frac{1}{4}{\theta _*^2}$,
where $|\theta_*|\leq 1$ and it is obtained from $\Lambda(l,r;i\theta_*)=0$ by
\begin{equation}\label{theta_*}
\cos\pi\theta_*=-lr+\sqrt{\left(1-l^2\right)\left(1-r^2\right)}~.
\end{equation}
The minimal allowed value $T_0=-\frac{1}{4}\,$  ($\,|\theta_*|=1$) is   
achieved for $l=r=-1$, and the corresponding Liouville field 
configuration $V_0=2m\sin\sigma$ is invariant under the $SL(2,\rr)$
conformal transformations. 

The general $V$-field is obtained by the action of the 2d conformal group on 
(\ref{V0})
\begin{equation}\label{conf-tr-V}
V_0(x,\bar x)\rightarrow V(x,\bar x)= \left(\xi'(x)\xi'(\bar
x)\right)^{-\frac{1}{2}}\,\,V_0(\xi(x),\xi(\bar x))~.
\end{equation}
The functions $\xi(x)$ here are monotonic $\xi'(x)>0$ and satisfy
the condition $\xi(x+2\pi)=\xi(x)+2\pi$. The variable $\tau_0$ in
(\ref{V0}) is absorbed by the zero mode of $\xi(x)$ and we find the
following parameterization of the Liouville field  in terms of
($T_0,$ $\,\xi(x)$) 
\begin{eqnarray}\label{V-xi}
V(x,\bar x)=\frac{m\left(\xi'(x)\xi'(\bar
x)\right)^{-\frac{1}{2}}}{p\,\sinh \pi p}
\Big (\sqrt{\Lambda(l,r;p)} \,e^{-\frac{p}{2}(\xi(x)+\xi(\bar x))}
+ \sqrt{\Lambda(l,r;p)} \,e^{\frac{p}{2}(\xi(x)+\xi(\bar x))}+~
\\~~~~~~~~
(l\,e^{-\pi p }+r )~
 e^{\frac{p}{2}(\xi(x)-\xi(\bar x))}+
 (l\,e^{\pi p}+r)
 e^{\frac{p}{2}(\xi(\bar x)-\xi(x))}
\Big ) ~,~~~p=2\sqrt {T_0}~.
\nonumber
\end{eqnarray}

 The action (\ref{S}) defines the canonical 2-form of the theory,
 which in terms of $\,(T_0 ,\,\xi(x))$ reads 
\begin{equation}\label{omega-FZZT}
\omega= \delta T_0 \wedge\int_0^{2\pi}dx\,\xi'(x)\delta\xi(x)+
T_0 \int_0^{2\pi}dx\,\delta\xi'(x)\wedge\delta\xi(x)+
\frac{1}{4}\int_0^{ 2\pi}dx\,
\frac{\delta\xi''(x)\wedge\delta\xi'(x)}{\xi'\,^2(x)}~.
\end{equation}
The energy-momentum tensor calculated from (\ref{V-xi}) is
given by 
\begin{equation}\label{T(x)-xi}
T(x)=\frac{\partial ^{\,2}_{xx}V}{V}~=
~T_0 \,\xi'\,^2(x)+\left(\frac{\xi''(x)}{2\xi'(x)}\right)^2-
\left(\frac{\xi''(x)}{2\xi'(x)}\right)'~,
\end{equation}
and the symplectic form (\ref{omega-FZZT}) provides
the Poisson brackets $\{T(x),\,\xi(y)\}=\xi(y)\,\delta(x-y)$ and  
$\{T(x),\,T_0\}=0$,
which means that $T(x)$ is  the generator of
the conformal transformations.

From (\ref{omega-FZZT}) one can also get the 
canonical Poisson bracket $\{T_0,\,\xi_0\}=1$, where $\xi_0$ is
the zero mode of $\xi(x)$: $\xi_0=\int_0^{2\pi}dx [\xi(x)-x]$.
If $T_0$ is negative, the
solution (\ref{V-xi}) oscillates in time and the zero mode $\xi_0$
becomes cyclic. Therefore in quantum theory
the negative values of $T_0$ have to be
quantized. The semi-classical calculation of these levels is in
accordance with the spectrum obtained by the bootstrap method
\cite{T, BRZ}.

The pair $(T_0, \xi(x))$ with $T_0>0$ can be mapped to the chiral
free-field
\begin{equation}\label{ff}
\phi(x)=\frac{p\xi(x)}{2}+\frac{1}{2}\log\xi'(x)-
\frac{1}{2}\log\frac{m\,\sqrt{\Lambda(l,r;p)}}{p\sinh \pi p}~,
\end{equation}
which can be expanded in the Fourier modes
\begin{equation}\label{phi-modes}
\phi(x)=\frac{1}{2}(q+px)+ {i}\sum_{n\neq 0}\frac{a_n}{2n}
\,e^{-inx}~,
\end{equation}
with $p>0$.  The Poisson brackets of these variables
are canonical
\begin{equation}\label{PB,a_n}
\{p,q\}=1,~~~~~~~~~\{a_n,a_m\}=i\,n\,\delta_{n+m,\,0}~,
\end{equation}
since the symplectic form (\ref{omega-FZZT}) becomes
canonical in terms of $\phi(x)$, and the energy-momentum tensor 
(\ref{T(x)-xi}) takes the 
free-field form with  an  improvement term
\begin{equation}\label{T(x)-phi}
T(x)=\phi\,'^{\,\,2}(x)-\phi''(x)~.
\end{equation}
The map (\ref{ff}) is invertible and by (\ref{V-xi}) we obtain the
free-field parameterization
\begin{equation}\label{V-phi}
V(x,\bar x)=e^{-[\phi(x)+\phi(\bar x)]}\left[1+m\,B_p\,A(x)+
m\,C_p\,A(\bar x)+m^2\, D_p\,A(x)\,A(\bar x)\right]~,
\end{equation}
with
\begin{equation}\label{b-p,c-p}
B_p=\frac{l\,e^{-\pi p}+r}{\sinh\pi p}~,~~~~
C_p=\frac{l\,e^{\pi p}+r}{\sinh\pi p}~,~~~~~~
D_p=\frac{ \Lambda(l,r;p)}{\sinh^2\pi p}~
\end{equation}
and
\begin{equation}\label{A}
A(x)=\frac{e^{-\pi p}}{2\sinh\pi p}\,\int_0^{2\pi}dy e^{2\phi(x+y)}~.
\end{equation}
Here $A(x)$ is the standard `screening charge': $A'(x)=e^{2\phi(x)}$. 
It is related to $A_p(x)$ used in \cite{DJ} by the $p$-dependent
factor $A_p(x)=2\sinh\pi p\, \,A(x)$.

Since $p>0$, we find that 
$\Phi(x,\bar x)=\phi(x)+\phi(\bar x)$ is the
$in$-field of the theory, and the term 
$m^2\, D_p\,A(x)\,A(\bar x)\,e^{-\Phi(x,\bar x)}$ in (\ref{V-phi}) 
corresponds to the $out$-field exponent.

It has to be mentioned that for $T_0<0$ the field $\phi(x)$ becomes complex
and the free-field parameterization fails. Peculiarities of this case will be
discussed in Section 5.

\subsection{Quantization}

We work in dimensionless variables and, to match with the notations
used in the bootstrap approach, from now we denote the measure of non
commutativity of the canonical variables by $2b^2$  
\begin{equation}\label{ccr}
[q,p]=i\hbar~,~~~~~[a_m,a_n]=\hbar\,m\,\delta_{m+n,0}~,~~~~~~~~~\hbar = 2b^2~.
\end{equation}
Note also that the Liouville field used in \cite{FZZ, T} is 
rescaled by a factor $b^{-1}$ relative to our $\varphi$. 

Quantization of the system in the free-field variables assumes a
realization of the  
commutation relations (\ref{ccr}) 
in the Fock space with a $p$-dependent vacuum $|p,\,0\rangle$ $(p>0)$,
defined in a standard way
\begin{equation}\label{p-vacuum}
a_n|p,\,0\rangle=0~,~~~~~~~~~~  n>0~.
\end{equation} 
We use the same notations for classical and corresponding quantum
expressions, which, in general, are deformed in a consistent way to
satisfy the principles of quantum theory.

The operators for chiral free-field exponentials $e^{2\alpha\phi(x)}$
are defined in the normal ordered form
\begin{equation}\label{e^phi}
e^{2\alpha\phi(x)}=e^{2\alpha\phi_{_{0}}(x)}\,
\,e^{2\alpha\phi_{_+}(x)}\,\,e^{2\alpha\phi_{_-}(x)}~,
\end{equation}
where
\begin{equation}\label{ff-decompos}
\phi_0(x)=\frac{q}{2}+\frac{px}{2},~~~\phi_+(x)= -{i}\sum_{n>
0}\frac{a_{-n}}{2n} \,e^{inx},~~~\phi_-(x)={i}\sum_{n>
0}\frac{a_n}{2n} \,e^{-inx}~,
\end{equation}
and the operator for the screening charge (\ref{A}) reads
\begin{equation}\label{^A}
A(x)=\frac{ \,e^{-\pi(p+ib^2\,)} }{2\sinh\pi(p+ib^2\,)}\,
\int_0^{2\pi}dy\,\, e^{2\phi(x+y)}~.
\end{equation}
The quantum energy-momentum tensor then becomes
\begin{equation}\label{T-phi^}
T(x)=\phi\,'^{\,\,2}(x)- (1+b^2) \,\phi''(x)~,
\end{equation}
with a normal ordered  $\phi\,'^{\,2}(x)$-term. The screening charge is 
a conformal scalar and the conformal
dimension of $e^{2\alpha\phi(x)}$ is (using  the notation
$Q=b^{-1}+b$ for the background charge)
\begin{equation}\label{Delta_alpha}
\Delta_{\alpha}=\alpha(1+b^2-\alpha b^2\,) =\alpha b(Q-\alpha b).
\end{equation}
The chiral operators $e^{2\alpha\phi(x)}$  and $A(x)$, together with
$p$-dependent coefficients, are the building blocks
for the vertex operators of the theory. They satisfy the following 
exchange relations
\begin{equation}\label{exch-p}
f_p\,e^{2\alpha\phi(x)}=e^{2\alpha\phi(x)}\,f_{p-2i\alpha\,b^2}~,
~~~~~~~~
f_p\,A(x)=A(x)\,f_{p-2ib^2}~.~~~~~~~
\end{equation}
\begin{equation}\label{exch-1}
e^{2\alpha\phi(x)}\,e^{2\beta\phi(y)}=
e^{2\beta\phi(y)}\,e^{2\alpha\phi(x)}\,
\,e^{-2i\pi\alpha\beta\,b^2\,\epsilon(x-y)}~,
~~~~~~~~~~~~~~~~~~~~
\end{equation}
\begin{equation}\label{exch-2}
A(x)\,e^{2\alpha\phi(y)}=
e^{2\alpha\phi(y)}\left[A(x)
\,e^{-2i\pi\alpha\,b^2\,\epsilon(x-y)}
+A(y)\,\frac{i\,\sin 2\pi\alpha b^2\,e^{\pi\left
(p-2i\alpha b^2-ib^2\right)\epsilon(x-y)}}
{\sinh\pi\left(p-2i\alpha b^2-ib^2\right)}\right]~,
\end{equation}
\begin{eqnarray}\label{A-A^}
A(x)\,A(y)e^{i\pi b^2\,\epsilon(x-y)}=A(y)\,A(x)\,
e^{-i\pi b^2\,\epsilon(x-y)}+
~~~~~~~~~~~~~~~~~~~~~~~~~~~~~~~~~~~~~~~~~~~~~
\\ \nonumber
\frac{i\sin\pi b^2}{\sinh\pi(p+2ib^2)}
\left(
e^{\pi\left(p+2ib^2\right)\epsilon(x-y)}\,A^2(y)-
e^{-\pi\left(p+2ib^2\right)\epsilon(x-y)}
\,A^2(x)
\right)~,~~~
\end{eqnarray}
where $\epsilon (x)$ is the stair-step function with $\epsilon (x)=1$ in
$0<x<2\pi$ 
and $\epsilon (x+2\pi)=\epsilon (x)+2$.
Eq. (\ref{exch-p}) and (\ref{exch-1}) follow  directly from
the canonical commutation relation (\ref{ccr}). 
The derivation of (\ref{exch-2}) for
$\alpha={1}/{2}$ one can find in \cite{DJ}. 
The method presented there is based on identities satisfied by
the stair-step function $\epsilon(x)$. That scheme can be easily 
generalized for arbitrary $\alpha$
and it reproduces (\ref{A-A^}) as well. These exchange relations
are helpful to verify the locality and Hermiticity properties
of the vertex operators and to develop their algebraic calculus. 

Another useful relation, which is usually used to bring the
operator products to the normal ordered form, reads
\begin{equation}\label{exch-3}
e^{2\alpha\phi_-(x)}\,e^{2\beta\phi_+(y)}=
e^{2\beta\phi_+(y)}\,e^{2\alpha\phi_-(x)}\,F^{2\alpha\beta b^2}(x-y)~,
\end{equation}
where $F$ is the function, which describes the short distance singularities
of the theory
\begin{equation}\label{f(z)}
F(x)=\exp\left(\sum_{n>0}\frac{e^{-inx}}{n}\right)=
\frac{1}{1-e^{-ix}}=
\frac{e^{\frac{i}{2}(x-\pi\epsilon(x))}}{2|\sin\frac{x}{2}|}~,
\end{equation}
with real $x$ understood as $x-i0$.

Applying this formula to the primary free-field exponential 
\begin{equation}\label{Psi_alpha}
\Psi_\alpha(x,\bar x)=
e^{-i\pi\alpha^2b^2}\,e^{2\alpha\phi(\bar
x)}\,e^{2\alpha\phi(x)}~
\end{equation}
(the phase
factor in front guarantees Hermiticity), 
one finds its normal ordered form 
\begin{equation}\label{N-Psi_alpha}
\Psi_\alpha(x,\bar x)=e^{4\alpha\phi_0(\tau)}
\,e^{2\alpha(\phi_+(\bar x)+\phi_+(x))}\,e^{2\alpha(\phi_-(\bar x)
+\phi_-(x))}\,\,(2\sin\sigma)^{-2\alpha^2 b^2}~.
\end{equation}
Obviously, this operator  is singular at the boundaries.

The simplest nontrivial vertex operator is $V(x,\bar x)$, 
which is constructed on the basis of (\ref{V-phi}).
It is given as a sum of four terms
\begin{equation}\label{^V^}
V(x,\bar x)=V_{in}(x,\bar x)+V_{_B}(x,\bar x)+V_{_C}(x,\bar
x)+V_{out}(x,\bar x)~,
\end{equation}
where $V_{in}$ and $V_{out}$ correspond to the $in$ and $out$-field
exponentials:
\begin{eqnarray}\label{^Vin}
&&V_{in}\,(x,\bar x)=e^{-\frac{i\pi b^2}{4}}\,e^{-\phi(\bar
x)}\,e^{-\phi(x)}~,
\\ \label{^Vb}
&&V_{_B}\,(\,x,\,\bar x)=m_b\,e^{-\frac{i\pi b^2}{4}}\,B_p\,e^{-\phi(\bar
x)}\,e^{-\phi(x)}\,A(x)~,
\\ \label{^Vc}
&&V_{_C}\,(\,x,\,\bar x)=m_b\,e^{-\frac{i\pi b^2}{4}}\,C_p\,A(\bar
x)\,e^{-\phi(\bar x)}\,e^{-\phi(x)}~,
\\ \label{^Vout}
&&V_{out}(x,\bar x)=m_b^2\,e^{-\frac{i\pi b^2}{4}}\,A(\bar
x)\,e^{-\phi(\bar x)}\,D_p\,e^{-\phi(x)}\,A(x)~,
\end{eqnarray}
with
\begin{equation}\label{B(p),C(p),D(p)}
B_p=\frac{l_b\,e^{-\pi(p-ib^2)}+r_b}{\sinh\pi p}~,~~~~~
C_p=\frac{l_b\,e^{\pi(p+ib^2)}+r_b}{\sinh\pi p}~,~~~~~
D_p=\frac{\Lambda(l_b,r_b;p)}{|\sinh\pi(p+ib^2)|^2}~,
\end{equation}
and $\Lambda(l_b,r_b;p)$ is given by (\ref{u}) with the replacement
$l\rightarrow l_b$, $r\rightarrow r_b$.
These $p$-dependent coefficients are the quantum analogs of
(\ref{b-p,c-p}) and the parameters ($l_b$, $r_b$, $m_b$) can be
interpreted as `renormalized' ($l$, $r$, $m$), respectively.
The guiding principles for the construction of the vertex operator 
(\ref{^V^}) are conformal symmetry, Hermiticity and
locality. These conditions fix $V(x,\bar x)$ up to three
free parameters ($l_b$, $r_b$, $m_b$) and it is natural to 
look for their exact
relation to the physical parameters ($l$, $r$, $m$)
\footnote{We keep the same letters as for the classical case.} 
 of the theory,
which are defined by the quantum version of the Liouville equation
\begin{equation}\label{L-eq}
\partial^2_{x\bar x}\,\varphi(x,\bar x)+m^2 e^{2\varphi(x,\bar x)}=0
\end{equation}
and the boundary conditions (\ref{FZZT-B}).    
In the next sub-section we will get  
\begin{equation}\label{m-m_b}
\frac{m_b}{m}=\frac{l_b}{l}=\frac{r_b}{r}=
\sqrt{\frac{\sin\pi b^2}{\pi b^2}}~.
\end{equation}
Note that the relation between $m_b$ and $m$ is the same as for the 
periodic Liouville theory.

\subsection{Bulk and  boundary vertex operators}

The bulk Liouville vertex operator $V_\alpha(x,\bar x)$ corresponding 
to the Liouville field exponential $e^{2\alpha\varphi(x,\bar x)}$
is given by
\begin{eqnarray}\label{V_alpha}
V_\alpha(x,\bar x)=\Psi_\alpha(x,\bar x)\,{\cal{V}}_\alpha(x,\bar x)~.
\end{eqnarray}
Here  ${\cal{V}}_\alpha(x,\bar x)$ is a conformal scalar, which 
can be expanded in a formal powers series in  $m_b$ and
the screening charges 
\begin{eqnarray}\label{V_alpha1}
{\cal{V}}_\alpha(x,\bar x)=\sum_{l=0}^{\infty}
m^l_b\sum_{k=0}^lA^{l-k}(\bar x)\,A^k(x)\,c^{l,k}_p(\alpha)=
1+m_b\left(A(\bar x)\,c^{1,0}_p(\alpha)+
A(x)\,c^{1,1}_p(\alpha)\right)~~~
\\ \nonumber
+\,m_b^2\left(A^2(\bar x)\,c^{2,0}_p(\alpha)
+A(\bar x)A(x)\,c^{2,1}_p(\alpha)+A^2(x)\,c^{2,2}_p(\alpha)\right)+\dots~,
~~~~
\end{eqnarray}
with $p$-dependent coefficients 
$c_p^{l,k}(\alpha)$, 
which we will calculate later.
Note that the quadratic term $A(x)A(\bar x)$ in (\ref{V_alpha})
is eliminated  by (\ref{A-A^}). Respectively,
in the general term one can choose the ordering where $A(\bar x)$ stands
left to $A(x)$.

The operator $V(x,\bar x)$ given by
(\ref{^V^})-(\ref{B(p),C(p),D(p)}), corresponds to $\alpha=-\frac{1}{2}$,
and by the exchange relations (\ref{exch-p})-(\ref{exch-2}) it takes the form 
(\ref{V_alpha})-(\ref{V_alpha1}) 
\begin{eqnarray}\label{^V1}
V(x,\bar x)=e^{-\frac{i\pi b^2}{4}}\,e^{-\phi(\bar
x)}\,e^{-\phi(x)}\Big[1+m_b\Big(A(\bar x)c^{1,0}_p+A(x)c^{1,1}_p\Big)+
~~~~~~~\\ \nonumber
~~~~~~~~~~~~~~~~~m_b^2\Big(A(\bar x)A(x)c^{2,1}_p+
A^2(x)c^{2,2}_p\Big)\Big]~.
\end{eqnarray}
Here we omit the index $\alpha=-\frac{1}{2}$ on both  sides
of the equation. The expansion in $m_b$ stops at the second order and besides 
$c_p^{2,0}=0$ we find 
\begin{eqnarray}\label{f-g,M-N}
c^{1,0}_p=\frac{l_b\,e^{\pi p}+r_b\,e^{-i\pi b^2}}
{\sinh\pi(p+ib^2\,)}
~,~~~~~~~~~~~
c^{1,1}_p=\frac{l_b\,e^{-\pi p}+r_b\,e^{i\pi b^2}}
{\sinh\pi(p+ib^2\,)}
~,~~~~~~~~~~~~~~~~
~~~~~~~~~~~~~~~~ \\ \label{M-N}
c^{2,1}_p=\frac{\Lambda(l_b,r_b;p-ib^2)\,
\,e^{-i\pi b^2}}{\sinh\pi p\,\,\sinh\pi(p-ib^2)}~,~~
c^{2,2}_p=-\frac{i\sin\pi b^2\,\,\,e^{-\pi(p-2ib^2)}\,\Lambda(l_b,r_b;p-ib^2)}
{\sinh\pi(p-2ib^2)\,\sinh\pi(p-ib^2)
\,\sinh\pi p}
~.
\end{eqnarray}
For convenience, here we have chosen  the ordering, 
where the primary free-field
exponential stands to the left, the $p$-dependent coefficients to the 
right and the screening charge operators between them. 
Using the exchange relations  (\ref{exch-p})-(\ref{exch-2}),
this structure can be preserved in algebraic calculations of the vertex
operators (see Appendix A).

The coefficients of the expansion (\ref{V_alpha1}) can be obtained
from the equation
\begin{equation}\label{VV=VV}
V_\alpha(x,-x)\,V(y,-y)=V(y,-y)\,V_\alpha(x,-x)~,~~~~~~~~(0<y<x<\pi)~,      
\end{equation}
which corresponds to the vanishing equal time commutator between
the $V$ and $V_\alpha$ fields.
Expanding the left and right hand sides
of eq. (\ref{VV=VV}) in powers of $m_b$ and using the above mentioned
ordering of the screening charge operators, we get similar operator
structures on  both sides of the equation. This relates
the $p$-dependent coefficients to each other and they can be
obtained step by step. This scheme is described in Appendix A in more detail.

An alternative scheme is based on the calculation of 
$V_{-\frac{n}{2}}$
operators, which correspond to the renormalized powers of (\ref{^V1}).
This scheme is also given in Appendix A, and both schemes lead to
the same result.
Here we consider  the following
three coefficients only
\begin{equation}\label{g^alpha_p}
c^{1,0}_p(\alpha)=s(\alpha)\,\,
\frac{l_b\,e^{\pi(p-4i\alpha b^2-2ib^2)}+r_b\,e^{-i\pi b^2}}
{\sinh\pi(p-4i\alpha b^2-ib^2)}\,\,
e^{2i\pi\alpha b^2+i\pi b^2}~,~~~~~~~~~~~~
\end{equation}
\begin{equation}\label{f^alpha_p}
c^{1,1}_p(\alpha)=s(\alpha)\,\,
\frac{l_b\,e^{-\pi(p-4i\alpha b^2-2ib^2)}+r_b\,e^{i\pi b^2}}
{\sinh\pi(p-4i\alpha b^2-ib^2)}\,\,
e^{-2i\pi\alpha b^2-i\pi b^2}~,~~~~~~~~~~~~
\end{equation}
\begin{eqnarray}\label{N^alpha_p}
c^{2,1}_p(\alpha)=e^{-i\pi b^2}\Big[s(\alpha)\,\,
\frac{\sinh\pi(p-2i\alpha b^2-2ib^2)}{\sinh\pi(p-4i\alpha b^2-2ib^2)}
~~~~~~~~~~~~~~~~~~~~~~~~~~~~~~~~~~\\ \nonumber 
~~~~~~~~~~~~~~~~~~~~~~~~~~~+s^2(\alpha)\,\,
\frac{l_b^2+r_b^2+2l_b^2\,r_b^2\cosh\pi(p-4i\alpha b^2-
3ib^2\,)}{\sinh\pi(p-4i\alpha b^2-2ib^2)\,
\sinh\pi(p-4i\alpha b^2-
3ib^2)}\Big]~,
\end{eqnarray}
where
\begin{equation}\label{s_alpha}
s(\alpha)=-\frac{\sin2\pi\alpha b^2}{\sin\pi b^2}~.
\end{equation}
These three coefficients are sufficient to establish 
the relations (\ref{m-m_b}).

The Liouville field operator is defined by 
\begin{eqnarray}\label{varphi}
\varphi(x,\bar x)=\frac{1}{2}~\partial_\alpha V_\alpha(x,\bar x)|_{\alpha=0}~,
\end{eqnarray}
and the operator Liouville equation reads
 \begin{equation}\label{^L-eq}
\partial^2_{x\bar x}\,\varphi(x,\bar x)+m^2 V_1(x,\bar x)=0~,
\end{equation}
where $V_1(x,\bar x)$ is the vertex operator for $\alpha=1$, 
with the conformal
dimension equal to one.

The expansion of the Liouville field in powers of $m_b$ by (\ref{V_alpha1})
leads to 
\begin{eqnarray}\label{varphi=}
\varphi(x,\bar x)=\phi(x)+\phi(\bar x)+
\frac{m_b}{2}\Big(A(x)\partial_\alpha c^{1,1}_p(0)+
A(\bar x)\partial_\alpha c^{1,0}_p(0)\Big)+\\ \nonumber
\frac{m_b^2}{2}\,A(\bar x)A(x)
\partial_\alpha c^{2,1}_p(0)
+\dots~.
\end{eqnarray}
Comparing then the mass square terms in (\ref{^L-eq}) and using that
$A'(x)=e^{2\phi(x)}$ (see (\ref{A})) we find
\begin{equation}\label{m=m}
\frac{m_b^2}{2}\,\,\,\partial_\alpha c^{2,1}_p(0)=-m^2
\,\,e^{-i\pi b^2}~,
\end{equation}
which by (\ref{N^alpha_p})-(\ref{s_alpha}) provides the ratio
${m_b}/{m}$ given just by
(\ref{m-m_b}).

Now we consider the quantum boundary conditions. They involve 
the boundary vertex operators, which
are defined as the boundary limit of the bulk operators
after dividing out the short distance singularity relative to the boundary  
\begin{equation}\label{V^l,V^r}
V_{2\alpha,\,l}(\tau)=\lim_{\sigma\rightarrow 0}V_\alpha(\tau,\sigma)
\,\left(2\sigma\right)^{2\alpha^2 b^2}~,~~~~~~~
V_{2\alpha,\,r}(\tau)=\lim_{\sigma\rightarrow \pi}V_\alpha(\tau,\sigma)
\,\left(2\pi-2\sigma\right)^{2\alpha^2 b^2}~.
\end{equation}
Note that here  the power of $2\sigma$ and $(2\pi -2\sigma)$ is just
$2\Delta _{\alpha}-\Delta _{2\alpha}$.    
The boundary behavior of the vertex operators $V_\alpha$ is governed by
$\Psi_\alpha$ and from
(\ref{N-Psi_alpha}) we obtain
\begin{eqnarray}\label{V^l=}
V_{\alpha,\,l}(\tau)&=&e^{4\alpha\phi(\tau)}
\left[1+m_b\,A(\tau)\left(c^{1,0}_p(\alpha)+
c^{1,1}_p(\alpha)\right)+\dots\right]~,\\[2mm]
V_{\alpha,r}(\tau)&=&e^{4\alpha\phi(\tau-\pi)}
e^{2\alpha\pi(p-i\alpha b^2)}
\left[1+m_b\,A(\tau-\pi)\left(c^{1,0}_p(\alpha)
+c^{1,1}_p(\alpha)\,e^{2\pi(p-ib^2)}\right)+
\dots\right].\nonumber
\end{eqnarray}
Here, for $V_{\alpha,\,r}(\tau)$, we have taken into account the monodromies
\begin{equation}\label{monodromy}
A(\tau+2\pi)=A(\tau)\,e^{2\pi(p-ib^2)}~,~~~~~~~
e^{4\alpha\phi_0(\tau)}=e^{4\alpha\phi_0(\tau-\pi)}\,
e^{2\alpha\pi(p-2i\alpha b^2)}~.
\end{equation}

The quantum version of the boundary conditions (\ref{FZZT-B}) assumes the form
\begin{equation}\label{d_sigma}
\partial_\sigma\varphi(\tau,\sigma)|_{\sigma=0}=
2ml\,V_{1,\,l}(\tau)~,~~~~~~
\partial_\sigma\varphi(\tau,\sigma)|_{\sigma=\pi}
=-2mr\,V_{1,\,r}(\tau)~.
\end{equation}
Since $A'(x)=e^{2\phi(x)}$, the first order mass terms of these equations
yield
\begin{eqnarray}\label{l,r=l,r}
&&\frac{m_b}{2}\left(\partial_\alpha c^{1,1}_p(0)-
\partial_\alpha c^{1,0}_p(0)\right)=2ml~,\\ \nonumber
&&\frac{m_b}{2}\left(\partial_\alpha c^{1,1}_p(0)\,
e^{2\pi(p-ib^2}-\partial_\alpha c^{1,0}_p(0)\right)=
2mr\,e^{\pi(p-2ib^2)}~,
\end{eqnarray}
and by (\ref{g^alpha_p})-(\ref{f^alpha_p}) we arrive at (\ref{m-m_b}).

\setcounter{equation}{0}

\section{Dynamics in the Morse potential}

The zero mode approximation of the periodic Liouville theory appears
rather useful for understanding the analytical properties of 
correlation functions of the theory \cite{Th,JW1}. This
approximation corresponds to $\sigma$-independent field
configurations, which are described by a particle dynamics in the
exponential potential. The particle model is exactly solvable
quantum mechanically \cite{D'H-J,BCGT}, and it allows a complete
description of the zero mode approximation. It is natural to look for a
similar approach to BLT. However, here the zero
mode sector corresponds to the $\sigma$-dependent solutions
(\ref{V0}). To proceed  nevertheless in this direction,  we make
the approximation on the level of  the initial action (\ref{S}) by putting
there $\varphi(\tau,\sigma)= y(\tau)$. In this way we find a mechanical
system with the Lagrangean
\begin{equation}\label{L}
L=\frac{1}{2}\,\dot y^2-\left[2
m^2\,e^{2y}+2m\lambda\,e^y\right]~,~~~~~~~~~~\lambda\equiv\frac{l+r}{\pi}~.
\end{equation}

This model describes the dynamics of a particle in the Morse potential
\begin{equation}\label{Morse}
{\cal M}(y)=2m^2\,e^{2y}+2m\lambda\,e^y~.
\end{equation}
As it is clear from Fig. 1, the case of negative $\lambda$ contains
both scattering and bound states like BLT. 
\begin{figure}[h]
\begin{center}
\includegraphics[height=4.8cm]{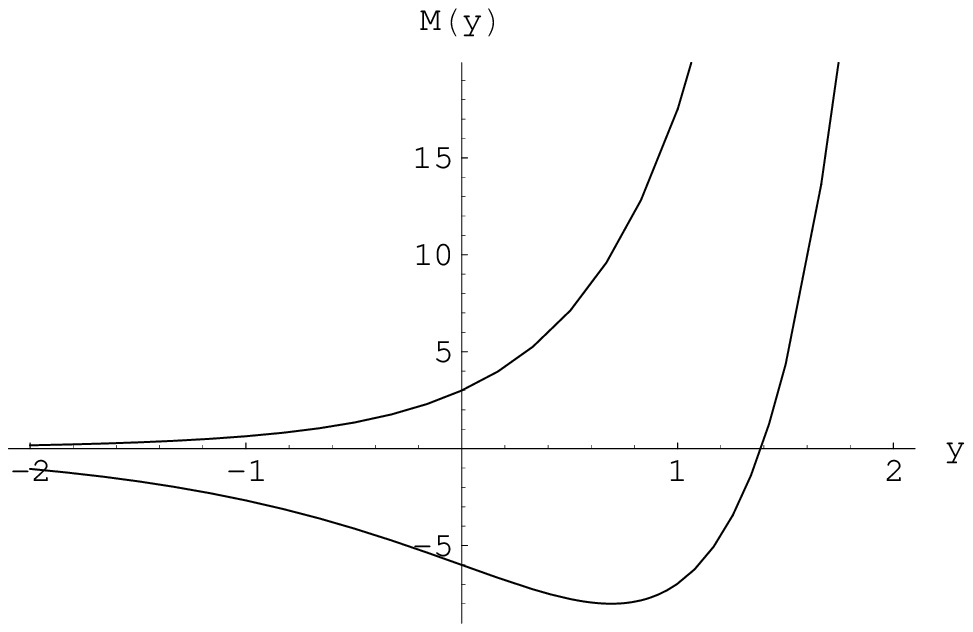}
\end{center}
\noindent {\bf Fig.1} {\it Typical form of the Morse potential ${\cal
 M}(y)$ for positive and negative $\lambda$.}
\end{figure}

For quantum mechanical systems the reflection amplitude and 
the discrete spectrum are usually obtained from the  
analysis of the asymptotic behavior of the solutions of the Schr\"odinger
equation. This equation for the Morse potential is exactly solvable
\cite{FL} (see Appendix B). However,   since a similar analysis of the
Schr\"odinger equation in field theory is rather problematic, we use this
toy model to introduce  an alternative method, 
based on the structure of the Heisenberg
operator $V(\tau)=e^{-y(\tau)}$. This method is then shown to reproduce the
exact results of Appendix B.

Classically $V(\tau)$ satisfies the equation
\begin{equation}\label{V-cl-eq}
V\ddot{V}-\dot{V}^2=4m^2+2m\lambda V~.
\end{equation}
Its general solution for positive energies $E=\frac{1}{2}\dot
y^2+U(y)>0$ can be written in terms of    asymptotic variables,
similarly to  (\ref{V-phi}) in BLT,
\begin{equation}\label{V-cl}
V(\tau)=e^{-(q+p\tau)}+e^{q+p\tau}\,F_\lambda(p)+G_\lambda(p)~,
\end{equation}
with $p=\sqrt{2E}>0$ and
\begin{equation}\label{F-G}
F_\lambda(p)=m^2\,\,\frac{\lambda^2+p^2}{p^4}~,~~~~~~
G_\lambda(p)=\frac{2m\lambda}{p^2}~.
\end{equation}
Negative energy solutions we get for imaginary $p=i\theta $. However, then
to stay with real $V$ in (\ref{V-cl}), the integration constant $q$
generically no longer can be chosen real. Instead it has to have the form
$q=- 1/2\log F_{\lambda}(i\theta)-i\beta$ with real $\beta$. Therefore, the
general solution of (\ref{V-cl-eq}) corresponding to negative
energies we represent as
\begin{equation}\label{V-cl-B}
V(\tau)=f_\lambda(\theta)\left(\,e^{i(\beta-\theta\tau)}+
e^{-i(\beta-\theta\tau)}\right)+g_\lambda(\theta)~,
\end{equation}
where $\theta=-\sqrt{2|E|}$ and $\,\beta$ is its canonically
conjugated cyclic variable ($\beta+2\pi\sim
\beta$) 
\beq\label{cyc}
\{\theta,e^{i\beta}\}=ie^{i\beta}~.
\eeq
In addition,
\begin{equation}\label{f-G}
f_\lambda(\theta)=\frac{m}{\theta^2}\,\,\sqrt{\lambda^2-\theta^2}~,~~~~~~
g_\lambda(\theta)=-\frac{2m\lambda}{\theta^2}~.
\end{equation}
The negative energy solutions obviously exist only for $\lambda<0$ 
and $\theta$ is bounded by 
\begin{equation}\label{lambda<}
-|\lambda| \leq \theta <0~. 
\end{equation}
The lower bound corresponds to the particle sitting at rest at the minimum of
the potential. Due to the
reflection symmetry ($\theta\mapsto -\theta, \beta\mapsto -\beta$)
of (\ref{V-cl-B}), we take $\theta$ only negative to avoid double
counting of solutions.

Eq. (\ref{V-cl}) defines a map from the asymptotic to the
interacting variables $(p,q)\mapsto \nolinebreak (p_y, y)$, which
is a canonical transformation $dp_y\,\wedge dy=dp\wedge dq$. The
parameters $p$ and $q$ in (\ref{V-cl}) correspond to the
$in$-momentum and $in$-coordinate, respectively. From (\ref{V-cl})
one can read off the $out$-variables as well $p_{out}=-p$,
$q_{out}=-q-\log F_\lambda(p)$. They are related to the
$in$-variables by the canonical transformation, which is a
combination of the $p$-dependent translation of $q$
\begin{equation}\label{q+a(p)}
(p,\,q)~\mapsto~(p,\,q+\log F_\lambda(p))~
\end{equation}
and the reflection $(p,\,q)\mapsto (-p,\,-q)$. The quantum analog of
the canonical transformation (\ref{q+a(p)}) is provided by the
reflection amplitude. To foresee its structure we write
(\ref{q+a(p)}) as an exponential action of Poisson brackets
generated by a function $\gamma_\lambda(p)$:
$(p,\,q)\,\mapsto\,e^{\{\gamma_\lambda(p)}\,(p,\,q)=
(p,\,q+\gamma_\lambda^{\,\prime}(p)).$
Then this function is defined up to an integration constant $c$
as a solution of the equation
\begin{eqnarray}\label{gamma(p)cl}
\label{gamma'(p)}
&&\partial_p\gamma_\lambda(p)=\log F_\lambda(p)~,\\
&&\gamma_\lambda(p)=c+2p+2p\log m+p\log(p^2+\lambda^2)-4p\log p
+2\lambda\,\arctan(p/\lambda)~.
\end{eqnarray}

Quantizing this mechanical system we use 
the canonical commutator $[y,p_y]=i\hbar$. 
The quantization of the system in the Schr\"odinger picture is summarized 
in Appendix B. There we also derive that the Heisenberg operator 
$V(\tau)$ satisfies as its dynamical equation
a deformed version of (\ref{V-cl-eq}) 
\begin{equation}\label{V-qu-eq}
\frac{1}{2}\,\left(V\ddot{V}+\ddot{V}V\right)
-\dot{V}^2+\frac{\hbar^2}{2}\,V^2=4m^2+2m\lambda V~.
\end{equation}
The quantum version of (\ref{V-cl}) we represent in the form
\begin{equation}\label{^V}
V(\tau)=e^{-q-p\tau}+e^{\frac{q}{2}}\,e^{p\tau}\,
F_\lambda\left(p\right)\,e^{\frac{q}{2}}+G_\lambda(p)~.
\end{equation}
This equation expresses the operator $V(\tau)$ in terms of ($p$,
$q$), which are the operators for the $in$-variables. In the
$p$-representation ($p>0$) $\,p\,$ acts as a multiplication and
$q=i\hbar\partial_p$. $F_\lambda(p)$ and $G_\lambda(p)$ in
(\ref{^V}) are the quantum analogs of (\ref{F-G}) and they could
have quantum deformations. Note that we have chosen the symmetric
ordering of $q$-exponentials in the second term of (\ref{^V}). To
fix the functions $F_\lambda(p)$ and $G_\lambda(p)$ in (\ref{^V}) we
use the dynamical equation (\ref{V-qu-eq}). With the help of
the exchange relations $e^{\pm q}\,f(p)=f(p\pm i\hbar)\,e^{\pm q}$
it  provides
\begin{equation}\label{^F-G}
F_\lambda(p)=m^2\,\,\frac{\lambda^2+p^2}
{p^2\,\left(p^2+\frac{\hbar^2}{4}\right)}~, ~~~~~~
G_\lambda(p)=\frac{2m\lambda}{p^2+\frac{\hbar^2}{4}}~.
\end{equation}

\subsection{The discrete spectrum in the Morse Potential}

The operator $\,V(\tau)\,$ in the sector of bound states is constructed
in a similar way, but now on the basis of (\ref{V-cl-B}).
The operator  $\theta$ has a discrete spectrum. 
Its  eigenvalues $\theta_n$
($n=0,1,..,N$) are negative and bounded below, 
$n=0$ corresponds to the ground state
and the eigenstates $|\theta_n\rangle$ are orthonormal. The
operators for the exponentials $e^{\pm i\beta}$ we denote by
$U_\pm$, which are the rising and lowering operators for $\theta$,
since they satisfy the commutation relations
$[\theta, U_\pm]=\pm \hbar\, U_\pm$ (see eq. (\ref{cyc})). 
A more general relation can be written in the exchange form
\begin{equation}\label{^U-pm-f}
f(\theta)\,U_\pm=U_\pm\,f(\theta\pm\hbar)~,
\end{equation}
for an arbitrary $f(\theta)$.
In particular, $e^{\mp i(\theta\mp\frac{\hbar}{2})\tau}\,U_\pm=
U_\pm\,e^{\mp i(\theta\pm\frac{\hbar}{2})\tau}.$ These operators are
the quantum   versions    of $e^{\pm i(\beta-\theta\tau)}$, since in the
scattering sector we have a similar correspondence, due to 
\begin{equation}\label{^e-pm}
e^{\pm(q+p\tau)}=e^{\pm q}\,e^{\pm(p\mp\frac{i\hbar}{2})\tau}=
e^{\pm(p\pm\frac{i\hbar}{2})\tau}\,e^{\pm q}~.
\end{equation}
After these remarks the operator $V(\tau)$ is represented in the form
\begin{equation}\label{^V-b}
V(\tau)=f_\lambda(\theta)\,e^{-i(\theta-\frac{\hbar}{2})\tau}\,U_+
+U_-\,e^{i(\theta-\frac{\hbar}{2})\tau}\,
f_\lambda(\theta)+g_\lambda(\theta)~,
\end{equation}
where  $f_\lambda(\theta)$ and $g_\lambda(\theta)$ are the quantum
versions of their classical counterparts in (\ref{V-cl-B}). 
These functions are fixed again by eq. (\ref{V-qu-eq}):
\begin{equation}\label{^f-g}
f_\lambda(\theta)=m\,\,\sqrt{\frac{\lambda^2-
(\theta-\frac{\hbar}{2})^2}
{\theta\,(\theta-\hbar)\left(\theta-\frac{\hbar}{2}\right)^2}}~,
~~~~~~
g_\lambda(\theta)=-\frac{2m\lambda}{\theta^2-\frac{\hbar^2}{4}}~.
\end{equation}
The operator (\ref{^V-b}) for $\tau=0$ acts  on the basis vectors
$|\theta_n\rangle$ ($n<N$) in the following way
\begin{equation}\label{V(0)psi}
V|\theta_n\rangle=f_\lambda(\theta_n+\hbar)\,|\theta_n+\hbar\rangle+
f_\lambda(\theta_n)\,|\theta_n-\hbar\rangle
+g_\lambda(\theta_n)\,|\theta_n\rangle ~.
\end{equation}
Since $\vert \theta_0\rangle $ is the ground state,  
$f_\lambda(\theta_0)=0$, and then by (\ref{^f-g})  
$\theta_0=\lambda+\frac{\hbar}{2}$. The function $f_\lambda(\theta)$ has no
other zeros for negative $\theta,$ and 
we get the equidistant spectrum
\begin{equation}\label{theta-n}
\theta_n=\lambda+\left(n+{1}/{2}\right)\hbar~,
\end{equation}
which coincides with the exact result given in Appendix B. The
eigenstates $|\theta_n\rangle$ correspond to the normalized wave
functions (\ref{psi-n}) and one can check that (\ref{V(0)psi}) holds
in the $y$-representation as well (see, for example, eq. (\ref{psi0-1})). 
This indicates the unitary equivalence of the $p$- and $y$-representations.
The numbers $f_\lambda(\theta_n)$ and $g_\lambda(\theta_n)$ are real,
therefore the operator $V$ is Hermitean. It is positive as well, only its
action is not defined on the highest level state $|\theta_N\rangle$,
like in the $y$-representation.

To establish the relation between the bound states and zeros of the reflection
amplitude,  we switch again to the scattering sector.

\subsection{Reflection amplitude in the Morse potential}

The first and the second terms in (\ref{^V}) are the 
$in$- and $out$-exponentials,
respectively. Therefore, they are related by the $S$-matrix
\begin{equation}\label{S-eq}
e^{-(q+p\tau)}\,{\cal S}={\cal S}\,e^{\frac{q}{2}}\,e^{p\tau}\,
F_\lambda\left(p\right)\,e^{\frac{q}{2}}~.
\end{equation}
The operator ${\cal S}$ can be represented in the form ${\cal
S}={\cal P}R_\lambda(p)$, where ${\cal P}$ denotes the parity
operation (${\cal P}p=-p {\cal P}$, ${\cal P}q=-q{\cal P}$) and
$R_\lambda(p)$ is the reflection amplitude, which 
by (\ref{S-eq}) satisfies  the
equation
\begin{equation}\label{R-eq}
R_\lambda(p+{i\hbar}/{2})=R_\lambda(p-i\hbar/2)\, F_\lambda(p)~.
\end{equation}
This type of equation for the reflection amplitude of the periodic Liouville
theory was considered in \cite{JW1,Pak}. Here we propose a more constructive
approach for its analysis. 

Writing the unitary operator $R_\lambda(p)$ in exponential form
$R_\lambda(p)=e^{-\frac{i}{\hbar}\,\gamma_\lambda(p)}$, from
(\ref{R-eq}) we get the quantum version of (\ref{gamma'(p)})
\begin{equation}\label{^gamma(p)}
\frac{2}{\hbar}\,\,{\sin\left(\frac{\hbar}{2}\,\partial_p\right)}
\,\,\gamma_\lambda(p) =\log F_\lambda\left(p\right) ~,
\end{equation}
where the function $F_\lambda(p)$ is given by (\ref{^F-G}). So it is already
deformed with respect to (\ref{F-G}).
The phase factor $\gamma_\lambda(p)$ is not defined uniquely from
(\ref{^gamma(p)}), since the operator
$\sin\left(\frac{\hbar}{2}\,\partial_p\right)$ has zero modes
$\psi_n(p)=e^{\frac{2\pi n}{\hbar}\,p}$. We first specify a special solution
of (\ref{^gamma(p)}), requiring analyticity in $\hbar$. For
this purpose we write (\ref{^gamma(p)}) just in the form
(\ref{gamma'(p)}) with a deformed (in powers of $\hbar$) right hand
side
\begin{equation}\label{^O}
\partial_p\gamma_\lambda(p) =\hat O_\hbar \,\,\log
F_\lambda\left(p\right)~,
\end{equation}
with
\begin{equation}\label{^O1}
~~~~~~~~~~~~~\hat
O_\hbar=\frac{\frac{\hbar}{2}\partial_p}
{\sin\left(\frac{\hbar}{2}\partial_p\right)}=
1+\frac{\hbar^2}{4\cdot3!}\,\partial_p^2+\dots~.
\end{equation}
To calculate the deformed function we represent $\log
F_\lambda\left(p\right)$ as a Fourier type integral, where the
integrand is given by the eigenfunctions of the operator $\hat
O_\hbar$. 
 According to (\ref{^F-G})
\begin{equation}\label{logF}
\log F_\lambda(p)= \log m^2+\log{\left(p^2+\lambda^2\right)}-
\log{p^2}-\log\left(p^2+\hbar^2/4\right)~,
\end{equation}
and using the integral representation (\ref{log=}) (see Appendix D), 
we find
\begin{equation}\label{logF-int}
\log F_\lambda(p)= \log m^2+\int_0^\infty
dt\,\left[A(t)\,e^{ipt}+B(t)\,e^{-ipt}+C(t)\right]~,
\end{equation}
with
\begin{equation}\label{A(t)-C(t)}
A(t)=B(t)=\frac{1+e^{-\frac{1}{2}\,\hbar
t}-e^{-|\lambda|t}}{t}~,~~~~~~~~C(t)=-\frac{2e^{-t}}{t}~.
\end{equation}
Replacing the action of the operator $\hat O_\hbar$ on 
(\ref{logF-int}) by its eigenvalues,
from (\ref{^O})-(\ref{^O1}) we obtain 
\begin{equation}\label{gamma=}
\gamma_\lambda^{\,_I}(p)=c+p\log m^2+\int_0^\infty dt\,\left[\tilde
A(t)\,e^{ipt}+\tilde B(t)\,e^{-ipt}+pC(t)\right]~,~~~~~~
\end{equation}
where the `tilded'
functions are
\begin{equation}\label{A(t),B(t)->}
\tilde A(t)=\frac{\hbar}{2i\,\sinh\left(\frac{\hbar
t}{2}\right)}\,\,A(t)~,~~~~~~~~~\tilde B(t)=-\frac{\hbar}{2i\,
\sinh\left(\frac{\hbar t}{2}\right)}\,\,B(t)~,
\end{equation}
$c$ is an integration constant and
we have used the index $I$ to distinguish this special
solution of (\ref{^gamma(p)}).
We choose $c=\pi\hbar$, 
providing $e^{-\frac{i}{\hbar}\gamma_\lambda^{\,_I}(0)}=-1$.
The integral in (\ref{gamma=}) splits into a sum of the three following terms
\begin{eqnarray}\label{I-1}
I_1&=&i\hbar\int_0^\infty \frac{dt}{t}\left[
\frac{e^{-(|\lambda|+\frac{\hbar}{2}-ip)t}-
e^{-(|\lambda|+\frac{\hbar}{2}+ip)t}}{1-e^{-\hbar t}}
-\frac{2ip}{\hbar}\,e^{-t}\right]~, \\ \label{I-2}
I_2&=&i\hbar\int_0^\infty \frac{dt}{t}\left( \frac{e^{-ipt}-
e^{ipt}}{1-e^{-\frac{\hbar t}{2}}}
+\frac{4ip}{\hbar}\,e^{-t}\right)~,\\ \label{I-3}
I_3&=&i\hbar\int_0^\infty \frac{dt}{t}\left(e^{ipt}-
e^{-ipt}\right)=-\pi\hbar~.
\end{eqnarray}
The integrals $I_1$ and $I_2$ take a more familiar form if
we rescale the integration variable in (\ref{I-1}) and
(\ref{I-2}) by $\hbar t\mapsto t$ and $\hbar t\mapsto 2t$, respectively. 
Then, due to (\ref{log(a)})
and (\ref{log-Gamma=}), eq. (\ref{gamma=}) becomes
\begin{equation}\label{^gamma(p)=}
\gamma_\lambda^{\,_I}(p)=2p\log\left(
\frac{4m}{\hbar}\right)+i\hbar\,\log\,
\frac{\Gamma\left(\frac{2ip}{\hbar}\right)}
{\Gamma\left(-\frac{2ip}{\hbar}\right)}\,
\frac{\Gamma\left(\frac{1}{2}+\frac{|\lambda|-ip}{\hbar}\right)}
{\Gamma\left(\frac{1}{2}+\frac{|\lambda|+ip}{\hbar}\right)}~.
\end{equation}
It is easy to check that the corresponding reflection amplitude
$R_\lambda^{\,_I}(p)=e^{-\frac{i}{\hbar}\,\gamma_\lambda^{\,_I}(p)}$ satisfies the initial
equation (\ref{R-eq}). Note that $F_\lambda(p)$ in this equation is an even
function of $\lambda$. Therefore both $R_\lambda(p)$ and $R_{-\lambda}(p)$
satisfy this equation, though they have to be different. 
To specify $R_\lambda(p)$
we make for the time being two assumptions: 

a) The phase $\gamma_\lambda^{\,_I}(p)$ given by (\ref{^gamma(p)=}) 
corresponds to
the phase of the reflection amplitude $R_\lambda(p)$ for $\lambda>0$.

b) The reflection amplitude $R_\lambda(p)$ is analytic in $\lambda$.

From these assumptions follows that, in general, 
we have again (\ref{^gamma(p)=}), 
but using $\lambda$ instead of $|\lambda|$. In this way we obtain 
\begin{equation}\label{R^s}
R_\lambda(p)=e^{-\frac{i}{\hbar}\,\gamma_\lambda(p)}=
 \left(\frac{4m}{\hbar}\right)^{-\frac{2ip}{\hbar}}~\,
\frac{\Gamma\left(\frac{2ip}{\hbar}\right)}
{\Gamma\left(-\frac{2ip}{\hbar}\right)}\,
\frac{\Gamma\left(\frac{1}{2}+\frac{\lambda-ip}{\hbar}\right)}
{\Gamma\left(\frac{1}{2}+\frac{\lambda+ip}{\hbar}\right)}~,
\end{equation}
which is just the reflection amplitude (\ref{Rk}) derived 
in the Schr\"odinger picture. Obviously this
$R_\lambda(p)$  also satisfies 
eq. (\ref{R-eq}). Considering $\gamma_\lambda(p)$ 
corresponding to (\ref{R^s}), and
using (\ref{Gamma-Gamma=ch}), we find 
\begin{equation}\label{gamma-gamma=}
\gamma_{\lambda}(p)-
\gamma_{-\lambda}(p)=2\pi\lambda+i\hbar
\,\log \left(\frac
{1+e^{\frac{2i\pi\lambda}{\hbar}}\,e^{-\frac{2\pi p}{\hbar}}}
{1+e^{-\frac{2i\pi\lambda}{\hbar}} \,e^{-\frac{2\pi
p}{\hbar}}}\right)~.
\end{equation}
Thus, $\gamma_{\lambda}(p)$ and $\gamma_{-\lambda}(p)$ differ from each other
by a term, which can be expanded in powers 
of $e^{-\frac{2\pi p}{\hbar}}$. 
Such  non-analytic terms are just the zero modes of the differential
operator in (\ref{^gamma(p)}), hence their presence is not surprising. 

Applying Stirling's formula (\ref{stirling}) to $\gamma_\lambda(p)$, defined
from (\ref{R^s}), 
we get the expansion in powers of $\hbar$
\begin{equation}\label{stirling-gamma}
\gamma_\lambda(p)\sim\gamma_{\lambda,\,_{0}}(p)+
\sum_{n\geq1}\, \gamma_{\lambda,\,n}(p)\, \hbar ^n~,
\end{equation}
where $\gamma_{\lambda,\,_{0}}(p)$ coincides with the classical solution
(\ref{gamma(p)cl}) for $c=0$, and the other expansion coefficients
are even functions of $\lambda$ as well. As a result,
$\gamma_\lambda(p)$ and 
$\gamma_{-\lambda}(p)$ have the same expansions in powers of $\hbar$, 
which can also be seen from
(\ref{gamma-gamma=}).
This sum is asymptotic, but it is Borel summable, 
and its Borel sum reconstructs the
function (\ref{^gamma(p)=}). A direct check of this statement
can be based on formula  (\ref{log-Gamma=1}), which represents $\log\Gamma(z)
-\log((z/e)^z\sqrt{2\pi/z})$ as a Laplace transform.

Three remarks are in order concerning our assumptions:

a) The calculation of $\gamma_\lambda^I(p)$ by the integral
representation corresponds to a spectral
decomposition with  a continuous spectrum. Since the case $\lambda>0$ 
does not contain a discrete spectrum
it is natural that $\gamma^{_I}_{\lambda }(p)$
reproduces the correct answer for $\lambda>0$.

b) The term $2m^2\,e^{2y}$ of the Morse potential is dominant in the reflection
wall and $2\lambda m\,e^{y}$ can be considered as a perturbation.
Therefore, the analyticity in $\lambda$ is natural as well.

c) In addition, the comparison with the exact reflection amplitude (\ref{R^s})
gives an independent proof that the assumptions are fully justified for the
Morse potential.
\vspace{2mm}

Finally, we can discuss how  for $\lambda <-\frac{\hbar}{2}$,  
the discrete spectrum
(\ref{theta-n}) manifests itself in the reflection amplitude
$R_\lambda(p)=e^{-\frac{i}{\hbar}
\gamma_\lambda(p))}$.  This function
has zeros at $p=i\theta_n$ ($n\geq 0$),
where $\theta_n=\lambda +n\hbar+\frac{\hbar}{2}$ 
are just the eigenstates 
(\ref{theta-n}).  
For  $\lambda \geq -\frac{\hbar}{2}$ there are no bound states, since
these zeros of $R_\lambda(p)$  now are on the positive imaginary axis.

\subsection{Correlation functions in the Morse potential}

At the end of this section we analyze the   `ground state'  matrix elements
of the operators $e^{2\alpha\,y}$, as a preparation for field-theoretical
calculation of correlation functions.   First we start with the
straightforward
calculations in the $y$-representation. The aim is
to develop an equivalent method in the $p$-representation, with a possible
generalization to BLT.

The `ket'  vectors $|p\rangle$ are the eigenstates 
of the Hamiltonian $H=p^2$.
We choose a normalization,
which in the $y$-representation  implies    the following
 asymptotic behavior at $y\rightarrow -\infty$
\begin{equation}\label{y-p}
\langle y|\,p\rangle\sim e^{\frac{i}{\hbar}\,py}+
R_\lambda(p)\,e^{-\frac{i}{\hbar}\,py}~.
\end{equation}
The wave functions $\langle y|\,p\rangle$ satisfy the Schr\"odinger
equation (\ref{Schr}) and  also the relation
\begin{equation}\label{<y|p>}
e^{-y}\langle y|\,p\rangle=\langle y|\,p+i\hbar\rangle
+F_\lambda(p-i\hbar/2)\langle y|\,p-i\hbar\rangle+
G_\lambda(p)\langle y|\,p\rangle~,
\end{equation}
which can be obtained from (\ref{^V}) at $\tau=0$, projecting it
between the states $\langle y|$ and $|p\rangle$.

The  analytical continuation of $\langle y|\,p\rangle$ in $p$,
at $p=i\theta_0=i(\lambda +\hbar/2)$   we denote by 
$\langle y|\,i\theta_0\rangle_{\nn}$. 
The state $|\,i\theta_0\rangle_{\nn}$ is a true ground state for $\lambda
<-\hbar/2$. We call it `ground state' for $\lambda\geq -\hbar/2$ as well,
though it is not a state in standard quantum mechanical sense.
A remarkable property of the `wave function' 
$\langle y|\,i\theta_0\rangle_{\nn}$ is its simple form, which follows from 
vanishing of the reflection amplitude $R_\lambda(p)$ at $p=i\theta_0$.
In some sense $|i\theta_0\rangle_{\nn}$ is an analog of 
the $SL(2,\rr)$ invariant
vacuum of Liouville theory and we  use it to calculate mean values of 
the operator $e^{2\alpha y}$.
Due to (\ref{Psi})  $\langle y|\,i\theta_0\rangle_{\nn}$ reduces to 
\begin{equation}\label{|lambda>}
\langle y|\,i\theta_0\rangle_{\nn}=e^{-\left(\frac{\lambda}{\hbar}+\frac{1}{2}\right)y}
\,e^{-\frac{2m}{\hbar}\,e^y}~.
\end{equation}
For $\lambda \geq -\hbar/2\,$ 
this function is not
normalizable and for 
$\lambda < -\hbar/2$ we get 
\begin{equation}\label{|lambda>=}
|\,i\theta_0\rangle_{\nn}=\left(\frac{4m}{\hbar}\right)
^{\frac{\lambda}{\hbar}+\frac{1}{2}}\,\sqrt{\Gamma
\left(-1-\frac{2\lambda}{\hbar}\right)}
~ |\theta_0\rangle~, 
\end{equation}
where  $|\theta_0\rangle $ is the corresponding normalized state with unit
norm.

The matrix element $\langle p'|e^{2\alpha\,y}|p\rangle$ 
is finite for real $p$, $p'$ and $\alpha >0$ . 
The calculation of $\langle p'|e^{2\alpha\,y}|p\rangle$ 
in the $y$-representation by (\ref{Psi}) leads to a rather
complicated answer in terms of hypergeometric functions \cite{GR}.
 The answer is simplified for    the state (\ref{|lambda>})
\begin{equation}\label{<0|Valpha|0>}
_{\nn}\langle i\theta_0|e^{2\alpha\,y}|i\theta_0\rangle_{\nn}=
\int_{-\infty}^\infty dy\,\, e^{\left(2\alpha-\frac{2\lambda}{\hbar}-1\right)y}
\,e^{-\frac{4m}{\hbar}\,e^y}=
\left(\frac{4m}{\hbar}\right)
^{\frac{2\lambda}{\hbar}-
2\alpha+1}\,\Gamma(2\alpha-\frac{2\lambda}{\hbar}-1)~.
\end{equation}
Note that      the integral in (\ref{<0|Valpha|0>})   
is finite even for non-normalizable 
$|i\theta_0\rangle_{\nn}$, if $\alpha$ 
is large enough: $\alpha>\frac{\lambda}{\hbar}+\frac{1}{2}$. 
 Then, the matrix element 
$_{\nn}\langle i\theta_0|e^{2\alpha\,y}|i\theta_0\rangle_{\nn}$ 
for generic $\alpha$ is understood as an 
analytical continuation of (\ref{<0|Valpha|0>}) in $\alpha$.

The correlation functions defined by the Heisenberg operators
 \begin{equation}\label{<0|V_alpha(t)|0>}
U_\alpha~=~ _{\nn}\langle i\theta_0|e^{2\alpha y(\tau)}|i\theta_0\rangle_{\nn}
\end{equation}
are time independent and, therefore, they 
  are given by  (\ref{<0|Valpha|0>}). 
  From this equation
follows that $4m^2\,U_1+2m\lambda\,U_{\frac{1}{2}}=0$, which is 
in accordance with the operator dynamical equation 
\begin{equation}\label{y"} 
\ddot {y}(\tau)+4m^2\,e^{2y(\tau)}+2m\lambda\,e^{y(\tau)}=0~,
\end{equation}
  even for non normalizable $|i\theta_0\rangle_{\nn}$.

The analytical continuation of (\ref{<0|Valpha|0>}) in $\alpha$, at 
$\alpha=-\frac{1}{2}$ yields
\begin{equation}\label{<i0|V|i0>}
_{\nn}\langle i\theta_0|V|i\theta_0\rangle_{\nn}
=\left(\frac{4m}{\hbar}\right)
^{\frac{2\lambda}{\hbar}+
2}\,\Gamma\left(-2-\frac{2\lambda}{\hbar}\right)=
\left(\frac{4m}{\hbar}\right)
^{1+\frac{2\theta_0}{\hbar}}\,\Gamma\left(-1-\frac{2\theta_0}{\hbar}\right)~.
\end{equation}
   
  The same matrix element for a normalizable $|i\theta_0\rangle_{\nn}$ can
also be calculated by the operator (\ref{^V-b}). Indeed,  from
(\ref{^V-b})-(\ref{V(0)psi})
we simply get 
\begin{equation}\label{<0|V|0>}
\langle \theta_0|V|\theta_0\rangle =
-\frac{2m}{\lambda+\hbar}~,
\end{equation} 
which together with
(\ref{|lambda>=}) reproduces (\ref{<i0|V|i0>}). 
This equation for positive $\lambda\,$ 
now can be treated as an
analytical continuation in $\lambda$ from the sector of bound states. 
Note that the norm of $|i\theta_0\rangle_{\nn}$
can also be calculated by the formula:
$|\,|\,i\theta_0\rangle_{\nn}\,|^2=
-i\hbar\,\,\partial_pR_\lambda(p)|_{p=i\lambda +i\hbar/2}$ 
(see (\ref{Psi-Psi})), which for the reflection amplitude (\ref{R^s}) 
 is equivalent to (\ref{|lambda>=}). 

Thus, the operator $V$ in the sector of bound states provides a simple way for 
the calculation of the correlation function (\ref{<i0|V|i0>}). 
Unfortunately this
scheme has no direct generalization to BLT, since the `ground state'
there is given by the $SL(2,\rr)$ invariant vacuum, which is normalizable
only for a certain value of the boundary parameters (see the next section) and
one can not make a continuation, like here in $\lambda$.

Now we consider an alternative scheme of derivation of (\ref{<i0|V|i0>})
using the operator (\ref{^V}). This scheme
is based on a regularization
of non-normalizable states $|i\theta_0\rangle_{\nn}$ in the $p$-representation.

Before starting    its discussion,   
a remark is in order: The operator $q=i\hbar\partial_p$   
is not self-adjoint in the Hilbert space $L^2(\rr_+)$ and the exponentials
$e^{\pm q}$ are not even Hermitean.  
This problem is typical for
Liouville theory, where the zero mode sector $(p,q)$ is   restricted to the
half-plane       $p>0$. On the other hand, the operator $V= e^{-y}$ 
is obviously self-adjoint in the $y$-representation. We assume   
that the $y$- and $p$-representations are unitary equivalent, but   
the    check     of this statement or a direct verification of   the     
self-adjointness of (\ref{^V}) needs         additional labor.  A proof   
of self-adjointness of the operator $V$ for the particle dynamics in the
Liouville potential ($\lambda =0$) one can find in \cite{JW1}. The proof is
based on a continuation of wave functions $\Psi(p)$ to the negative half-line,
and then on the full line the  Hermitean structure of $V$ becomes crucial.
This scheme can be easily generalized to the Morse potential. For this purpose
we consider the space of square-integrable 
functions
$\Psi(p)$ given by 
\begin{equation}\label{psi(p)=}
\Psi(p)=e_\lambda(p)\,\psi(p)~,
\end{equation} 
where $\psi(p)$ is an even 
$~(\psi(-p)=\psi(p))~$ holomorphic function and 
\begin{equation}\label{e(p)=}
e_\lambda(p)=\left(\frac{4m}{\hbar}\right)^{\frac{ip}{\hbar}}\,
\frac{\Gamma\left(\frac{1}{2}+\frac{\lambda+ip}{\hbar}\right)}
{\Gamma\left(\frac{2ip}{\hbar}\right)}~.
\end{equation}
This function satisfies the condition 
\begin{equation}\label{e(-p)/e(p)}
\frac{e_\lambda(-p)}{e_\lambda(p)}=R_\lambda(p)~,  
\end{equation}
and, therefore,  the continuation of (\ref{psi(p)=})
to the negative half-line has the following reflection property
\begin{equation}\label{psi(-p)=}
\Psi(-p)=R_\lambda(p)\,\Psi(p)~,
\end{equation} 
providing
\begin{equation}\label{int_0}
\int_0^\infty dp\,\, \Psi^*(p)\,V\Psi(p)=
\frac{1}{2}\,\int_{-\infty}^\infty dp\,\, \Psi^*(p)\,V\Psi(p)~,
\end{equation} 
which leads to the
self-adjointness  of the operator (\ref{^V}). 
\vspace{2mm}

Let us regularize the `ground state' $|i\theta_0\rangle$ by 
$e^{2\alpha y}|i\theta_0\rangle$, with a positive large enough parameter 
$\alpha$, and consider the corresponding wave function
in the $p$-representation
\begin{equation}\label{Psi_alpha(p)}
\Psi_\alpha(p)=\langle p|e^{2\alpha y}|i\theta_0\rangle~.
\end{equation} 
Using the integral (\ref{int(We)}) we find that $\Psi_\alpha(p)$
has the form (\ref{psi(p)=}) 
$\Psi_\alpha(p)=e_\lambda(p)\,\psi_{\alpha,\lambda}(p)$ with
\begin{equation}\label{psi_alpha(p)}
\psi_{\alpha,\lambda}(p)=
\left(\frac{4m}{\hbar}\right)^{-2\alpha+\frac{\lambda}{\hbar}
+\frac{1}{2}}\,\,\,
\frac{\Gamma\left(2\alpha-\frac{1}{2}-\frac{\lambda+ip}{\hbar}\right)
\Gamma\left(2\alpha-\frac{1}{2}-\frac{\lambda-ip}{\hbar}\right)}
{\Gamma(2\alpha)}~.
\end{equation}
We represent the matrix element $\langle\Psi_\alpha|V(0)|\Psi_\alpha\rangle$ 
as a sum $V_1+V_2$, where $V_1$ corresponds to the contribution
from the first two terms of (\ref{^V}) and $V_2$ reads
\begin{equation}\label{V_2}
V_2=\int_{-\infty}^{+\infty} \frac{dp}{2\pi\hbar}\,e(-p)\,e(p)\,
\psi^2_{\alpha,\lambda}(p)\,G_\lambda(p)~.
\end{equation} 
Note that the extension of the integration to 
the negative half-line is allowed, due to the reflection 
($p \leftrightarrow -p$) symmetry
of the integrand, and the integration measure here corresponds 
to our  normalization of $|p\rangle$-states.
The term $V_1$ is simplified with the help of (\ref{e(-p)/e(p)}) and 
(\ref{R-eq}) in the form
\begin{equation}\label{V_1}
V_1=\int_{-\infty}^{+\infty} \frac{dp}{\pi\hbar}\,\,\,e(-p)\,
\psi_{\alpha,\lambda}(p)\,e(p-i\hbar)\,
\psi_{\alpha,\lambda}(p-i\hbar)~, 
\end{equation} 
where the integration over the negative half-line now corresponds to
the second term (the $out$-term) of (\ref{^V}).
The integrals (\ref{V_2}) and (\ref{V_1}) are well defined due to 
the asymptotic behavior (\ref{Gamma(x+iy)}). 

  To remove the regularization (\ref{Psi_alpha(p)}) by sending
$\alpha\rightarrow 0$, one cannot immediately use the integral  
representations (\ref{V_2})-(\ref{V_1}), since they break down at 
$2\alpha=\frac{1}{2}+\frac{\lambda}{\hbar}$, due to singularities
of the integrands. Therefore, we first shift the contour of integration 
\begin{equation}\label{V_1,2->}
\int_{-\infty}^{+\infty} dp \rightarrow \int_{ia-\infty}^{ia+\infty} dp+
2\pi i\sum \mbox{Res}~, 
\end{equation} 
with $a$ taken from the interval
$(\frac{1}{2}+\frac{\lambda}{\hbar},~\frac{3}{2}+\frac{\lambda}{\hbar})$.
Now the integrals exist for all values of $\alpha$ down to zero
and disappear in the limit, since the  integrands tend to zero.     

The pole of the integrand in (\ref{V_2}) at 
$p=i(\lambda+\hbar/2)=i\theta_0$ provides the following
residue term  
\begin{eqnarray}\label{Res}
\frac{i}{\hbar}\,\,\mbox{Res} \,\,\Big[e(-p)\,e(p)\,
\psi^2_{\alpha,\lambda}(p)\,G_\lambda(p)\Big]|_{p=i\theta_0}=
\left(\frac{4m}{\hbar}\right)^{-4\alpha+\frac{2\,\theta_0}{\hbar}
}\,\,\frac{\Gamma^2\left(2\alpha-\frac{2\,\theta_0}{\hbar}\right)}
{\Gamma\left(-\frac{2\,\theta_0}{\hbar}\right)}
\frac{2m}{-\lambda-\hbar}~.
\end{eqnarray}
The other residue terms contain $\Gamma(2\alpha)$ in the denominator
and they vanish in the limit $\alpha\rightarrow 0$. As a result,
only  the term (\ref{Res}) survives in the limit $\alpha\rightarrow 0$
and we end up with (\ref{<i0|V|i0>}), which is given by 
\begin{equation}\label{R'}
_{\nn}\langle i\theta_0|V|i\theta_0\rangle_{\nn}=G(i\theta_0)\,
\left[-i\hbar\,\partial_pR_\lambda(p)\right]_{p=i\theta_0}~,
\end{equation}
even for a non-normalizable ground state $|i\theta_0\rangle_{\nn}$.
\vspace{2mm}

Summarizing this section we conclude that the structure of the operator 
$V$ in terms of the asymptotic variables
can be used for calculations of the discrete spectrum, the
reflection amplitude and the correlation function in a rather
simple way. We will apply these schemes to BLT in the next section.

\setcounter{equation}{0}

\section{The vacuum sector of BLT}

Let us consider the projection of the operator (\ref{^V^})
on the vacuum sector $\langle p',\,0|V|p,\,0\rangle$,
where $|p,\,0\rangle$ is the  $p$-dependent vacuum (\ref{p-vacuum}).
The matrix elements    $\langle p',\,0|V|p,\,0\rangle$ 
yield  an operator kernel in the
$p$-representation of the zero mode sector. 
We denote the corresponding operator 
by $\hat V$. Then, using eqs. (\ref{exch-3})-(\ref{f(z)})
to bring the operators (\ref{^Vin})-(\ref{^Vout}) to the
normal ordered form, we  obtain
\begin{equation}\label{<Vin>}
(2\sin\sigma)^{\frac{b^2}{2}}\,\hat V_{in}=e^{-(q+p\tau)}
~,~~~~~~~~~~~~~~~~~~~~~(2\sin\sigma)^{\frac{b^2}{2}}~\hat 
V_{out}=m_b^2\,e^{\frac{q}{2}}\,\,e^{p\tau}\,D_p\,I_p\,
e^{\frac{q}{2}} ~,~~~
\end{equation} 
\begin{equation}\label{<Vb>}
(2\sin\sigma)^{\frac{b^2}{2}}\,\hat V_{_B}=m_b\,B_p\,e^{(p-ib^2)
(\sigma-\pi)}\,J_p
~,~~~~~~(2\sin\sigma)^{\frac{b^2}{2}}\,~\hat V_{_C}=
m_b\,C_p\,
e^{-(p+ib^2)(\sigma+\pi)}
\,\bar J_p~,
\end{equation}
where $I_p$, $J_p$ and $\bar J_p$ contain 
the integrals provided by the screening charges
\begin{eqnarray}\label{Ip}
I_p=\frac{e^{-2\pi p}}{4\sinh^2\pi p}
\int_0^{2\pi}dy\int_0^{2\pi}d\bar y
~e^{p(y+\bar y)}\,\,\left(1-e^{iy}\right)^{b^2}\,\left(1-e^{-i\bar
y}\right)^{b^2}\times~~~~~
~~~~~~~~~~~~~~~~~~~~\\ \nonumber
\,\left(1-e^{2i\sigma}\,e^{iy}\right)^{b^2}\,
\left(1-e^{2i\sigma}\,e^{-i\bar y}\right)^{b^2}\,
\left(1-e^{2i\sigma}\,e^{i(y-\bar y)}\right)^{-2b^2}~,
\end{eqnarray}
\begin{equation}\label{Jp}
J_p=\frac{1}{2\sinh\pi(p-ib^2\,)}
\int_0^{2\pi}dy\,e^{(p-ib^2)y}\,
\left(1-e^{iy}\right)^{b^2}
\,\left(1-e^{2i\sigma}\,e^{iy}\right)^{b^2}~,~~~~~~~~~~~~~~~~~
\end{equation}
\begin{equation}\label{barJp}
\bar J_p=\frac{1}{2\sinh\pi(p+ib^2\,)}
\int_0^{2\pi}d\bar y\,
e^{(p+ib^2)\bar y}\left(1-e^{-i\bar
y}\right)^{b^2} \left(1-e^{2i\sigma}\,e^{-i\bar
y}\right)^{b^2}~.~~~~~~~~~~~~~~~
\end{equation}

$I_p$ from (\ref{Ip}) can     
be written with    (\ref{f(z)}) in the form
\begin{eqnarray}\label{Ip=R}
I_p(\sigma)=\frac{e^{-2\pi p}}{4\sinh ^2\pi p}\times 
~~~~~~~~~~~~~~~~~~~~~~~~~~~~~~~~~~~~~~~~~~~~~~~~~~~~~~~~~~~~~~~~~~~~~
~~~~~~~~\\
\int_0^{2\pi}dy\int_0^{2\pi}d\bar y~e^{p(y+\bar y)}\,\,
\frac{|4\sin\left(\frac{y}{2}\right)\,\sin\left(\frac{\bar y}{2}\right)\,
\sin\left(\frac{y}{2}+\sigma\right)\,\sin\left(\frac{\bar y}{2}-
\sigma\right)|^{b^2}}
{|\sin\left(\frac{y-\bar y}{2}+\sigma\right)|^{2b^2}}\,
e^{-i\pi b^2\,\alpha(y,\bar y,\sigma)}~,\nonumber
\end{eqnarray}
 where the phase of the last factor is given by
\begin{equation}\label{beta=}
\alpha(y,\bar y,\sigma )=\frac{1}{2}\,\epsilon(y+2\sigma)-\frac{1}{2}\,
\epsilon(\bar y-2\sigma)
-\epsilon(y-\bar y+2\sigma)~.
\end{equation}
This function takes only discrete values, 
$-1$, $0$ and $1$, as
they are indicated in  Fig. 2. The imaginary
\begin{figure}[h!]
\begin{center}
\includegraphics[height=4.8cm]{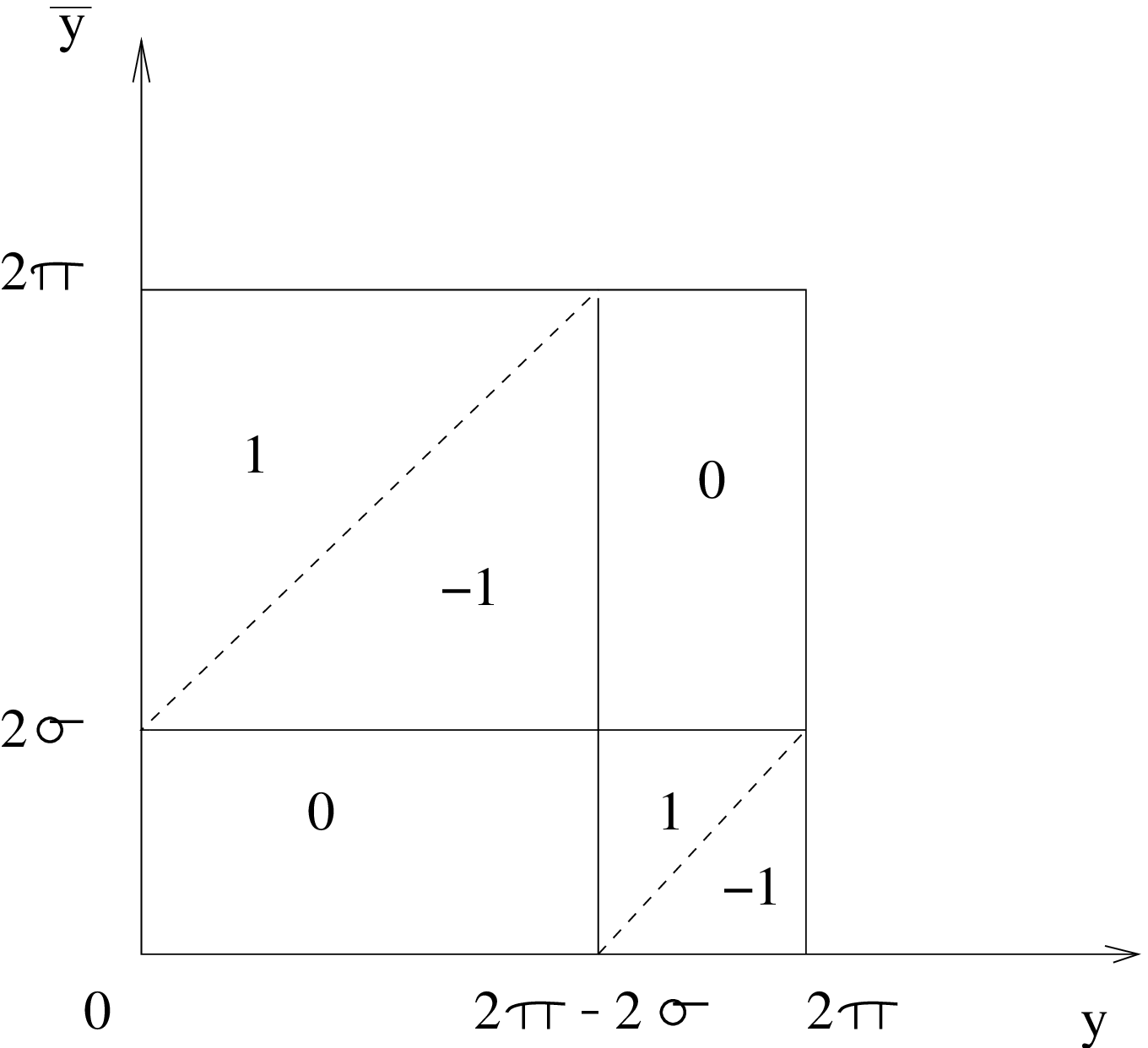}
\\[5mm]
 \end{center}
\noindent
{\bf Fig.2}~~{\it Values taken by the function $\alpha (y,\bar y,\sigma)$ 
in the  $(y,\bar y)$-plane.} 
\end{figure}
\noindent
part of   (\ref{Ip=R}) 
can arise only due to the integration
over the two small squares with the dash-line diagonals. But the modulus 
of the integrand is a symmetric function with respect to these diagonals,
and, since the constant phases differ only by their sign, the imaginary
contributions cancel each other.  
On the other hand, if we  look at $I_p$ as a function of 
the complex variable $\zeta
=e^{2i\sigma}$,  it is obviously holomorphic inside the unit circle. But since
$I_p$ is real on the unit circle, it is  $\zeta$-independent. 
At $\zeta=0$ the integral in (\ref{Ip}) splits into the product 
of two integrals of the type (\ref{int}) and, using (\ref{Gamma-Gamma=ch}), 
we find
\begin{equation}\label{Ip=}
I_p=\frac{\Gamma^{\,2}\left(1+b^2\right)\,
\Gamma\left(ip\right)\,
\Gamma\left(-ip\right)}
{\Gamma(1+b^2+ip)\,\Gamma(1+b^2-ip)}~.
\end{equation}
Expanding the integral (\ref{Jp}) in powers of   
$\zeta$: $J_p=\sum_{n\geq 0} j_n\,\zeta^n$,   
the coefficients $j_n$  are also obtained from (\ref{int})  
\begin{equation}\label{<bn}
j_n=\frac{\Gamma(n-b^2)}
{\sinh\pi(p-b^2)\,\Gamma(-b^2)
\,n!}\,\,
\frac{\pi\,\Gamma(1+b^2)\,e^{\,\pi\,(p+in-ib^2)}}
{\Gamma(1+ip-n+b^2)\Gamma(1-ip+n)}~.
\end{equation}
One observes that $j_n/j_0$ are the coefficients of the expansion for the
hypergeometric function. The calculation of $\bar J_p$ is similar
and from (\ref{<Vb>}) we get
\begin{equation}\label{<Vb>=}
(2\sin\sigma)^{\frac{b^2}{2}}\,\hat V_{_B}
=b_p\,\,e^{(p-ib^2)\sigma}
\,_2F_1(-b^2,-b^2-ip,1-ip,\zeta)~,
\end{equation}
\begin{equation}\label{<Vc>=}
(2\sin\sigma)^{\frac{b^2}{2}}\,\hat V_{_C} =
c_p\,e^{-(p+ib^2)\sigma}
\,_2F_1(-b^2,-b^2+ip,1+ip,\zeta)~,
\end{equation}
where $b_p=m_b\,B_p\,j_0\,e^{-\pi(p-ib^2)}\,$, 
$\,c_p=m_b\,C_p\,\bar j_0\,e^{-\pi(p+ib^2)}$, and
by (\ref{B(p),C(p),D(p)}) they become
\begin{equation}\label{bp}
b_p~=~c_{-p}~=~\frac{m_b}{\pi}\,\left(l_b\,e^{-\pi(p-ib^2)}+r_b\right)
\,\Gamma(1+b^2)\, \Gamma(ip)\,\Gamma(-ip-b^2) ~.
\end{equation} 
Thus, the operator $\hat V$   
has the form (\ref{^V})
\begin{equation}\label{<V>=}
\hat V~=~(2\sin \sigma )^{-\frac{b^2}{2}}\left (e^{-(q+p\tau)}+
e^{\frac{q}{2}}\,e^{p\tau}\,
F(p)\,e^{\frac{q}{2}}+G(p,\sigma)\right ) ~,
\end{equation}
with $F(p)=m_b^2\, D_{p}\,I_{p}$
and     $G(p,\sigma)$ given as a sum of
(\ref{<Vb>=}) and (\ref{<Vc>=}). 
The function $F(p)$ in  explicit form is obtained 
from (\ref{B(p),C(p),D(p)}), (\ref{Ip=}) and  (\ref{u})
\begin{equation}\label{F(p)=}
F(p)=\frac{m_b^2\,\,\Gamma^{\,2}(1+b^2)\,\Gamma(ip)\Gamma(-ip)
\,\Gamma(1-b^2+ip)\,\Gamma(1-b^2-ip)}
{\pi^2\left(p^2+b^4\right)}\,\,\Lambda(l_b,r_b;p)~,
\end{equation}
 while 
\bea
\label{G(p)=}
G(p,\sigma )&=&\frac{m_b}{\pi}\,\left(l_b\,e^{-\pi(p-ib^2)}+r_b\right)
\,\Gamma(1+b^2)\, \Gamma(ip)\,\Gamma(-ip-b^2)~e^{(p-ib^2)\sigma }~\nonumber\\
&\times &_2F_1(-b^2,-b^2-ip,1-ip,e^{2i\sigma})
~+~(p\leftrightarrow -p)~.
\eea
Note that both $F$ and $G$ are even functions of $p$ . Furthermore,    
$F$ is manifestly    real and the reality of $G(p,\sigma)$
follows from the analytical properties
of the hypergeometric functions (\ref{F(1/z)}).
This provides a Hermitean structure   for $\hat V$  and gives some consistency 
check for    the construction of $V$. 
The positivity of the operator   $\hat V$,   
which is also related to the
unitarity of the theory, puts certain restrictions
on the parameters $l_b$ and $r_b$ like in the classical case. 
  But,  due to the quantum deformations, 
the corresponding analysis becomes more complicated. Note that in the
limit $b\rightarrow 0$ one reproduces the classical expressions 
corresponding to (\ref{V-phi}) for vanishing Fourier modes $a_n=0$
\begin{equation}\label{F,G->hbar}
F(p)\rightarrow \frac{m^2\,\Lambda(l_b,r_b;p)}{p^2\sinh^2\pi p}~.
~~~~~~~G(p,\sigma)\rightarrow \frac{2m}{p\sinh\pi p}
\left[l\cosh p(\sigma-\pi)+r\cosh p\sigma\right]~.
\end{equation}
\vspace{2mm}

The boundary   parameters    of the classical theory are given by
the derivatives of the $V$-field at the boundaries (see (\ref{FZZT-B}))
\begin{equation}\label{partial V=}
\partial_\sigma V(\tau,\sigma)|_{\sigma=0}=-2ml~,~~~~~~
\partial_\sigma V(\tau,\sigma)|_{\sigma=\pi}=2mr~.
\end{equation}
To find the corresponding quantum relation note that by (\ref{<V>=})
\begin{equation}\label{partial<V>=}
\partial_\sigma\Big[(2\sin\sigma)^{\frac{b^2}{2}}\,
\hat V(\tau,\sigma)\Big]=
\partial_\sigma G(p,\sigma)~.
\end{equation}
The boundary behavior of $\partial_\sigma G(p,\sigma)$ can be obtained
from (\ref{F(1-z)})
\begin{eqnarray}\label{partialG=}
\partial_\sigma G(p,\sigma)|_{\sigma   \rightarrow    0}=-2m_b\,l_b\,\,
\frac{\Gamma(1+b^2)\Gamma(1-2b^2)}{\Gamma(1-b^2)}\,\,
(2\sigma)^{2b^2}(1+O(\sigma))~,~~~~~~~~~
\\ \nonumber
\partial_\sigma G(p,\sigma)|_{\sigma   \rightarrow     \pi}=2m_b\,r_b\,\,
\frac{\Gamma(1+b^2)\Gamma(1-2b^2)}{\Gamma(1-b^2)}\,\,
(2\pi-2\sigma)^{2b^2}
(1+O(\pi-\sigma))~.
\end{eqnarray}
Then, taking into account (\ref{m-m_b}), the quantum version of 
(\ref{partial V=}) becomes
\begin{eqnarray}\label{<partial V>=}
\lim_{\sigma\rightarrow 0} (2\sin\sigma)^{-2b^2}\,
\partial_\sigma\Big[(2\sin\sigma)^{\frac{b^2}{2}}\,  \hat V    (\tau,\sigma)
\Big]=
-2ml\,\,\frac{\Gamma(1-2b^2)}{\Gamma^2(1-b^2)}\,~,
\\ \nonumber
\lim_{\sigma\rightarrow \pi} (2\sin\sigma)^{-2b^2}\,
\partial_\sigma\Big[(2\sin\sigma)^{\frac{b^2}{2}}\,  \hat V   (\tau,\sigma)
\Big]\,\,=\,\,
2mr\,\,\frac{\Gamma(1-2b^2)}{\Gamma^2(1-b^2)}~.~
\end{eqnarray}
This result reflects part of the structure which one would expect from a 
boundary OPE for $V$ \cite{FZZ} at $\sigma\rightarrow 0$ 
\begin{equation}\label{OPE}
V(x,\bar x)=(2\sigma )^{\Delta _{-1}-2\Delta _{-1/2}}~V_{-1,l}(\tau)~+~(2\sigma
)^{-2\Delta _{-1/2}}~Z(b)~\mbox{\bf 1}~+~\dots ~.
\end{equation}
$V_{-1,l}(\tau)$  is defined in (\ref {V^l,V^r}). Projecting
(\ref{OPE}) to the vacuum sector, using (\ref{Delta_alpha}),  and acting with
$(2\sigma)^{-2b^2}\partial _{\sigma} (2\sigma) ^{\frac{b^2}{2}}$, a comparison
with (\ref{<partial V>=}) gives for the bulk-boundary structure constant
  $Z(b)$  
\beq\label{bulk-bound-str}
 Z(b)  ~=~ -\frac{2ml}{b^2}~\frac{\Gamma (-1-2b^2)}{\Gamma ^2 (-b^2)}~,
\eeq
which after identifying $\frac{ml}{\pi b^2}$ with $\mu _B$ (see below)
coincides with those in \cite{FZZ}.
\vspace{2mm}

Before starting calculations in the vacuum sector, let us make few remarks
concerning the Virasoro generators and the $SL(2,\rr)$
invariant vacuum state. The Virasoro generators are the Fourier modes of the
energy momentum tensor (\ref{T-phi^}) with an appropriate normalization
and a constant shift in $L_0$, to match the Virasoro algebra in 
the standard form
\begin{equation}\label{virasoro}
[L_n,L_m]=(n-m)L_{n+m}+
\frac{c}{12}(n^3-n)\delta_{n+m,0}~.
\end{equation}
Equation
 (\ref{T-phi^}) defines the following normalization and the shift 
\begin{equation}\label{l_n}
L_n=\frac{1}{2\pi b^2}\,\,\int_0^{2\pi}dx\, e^{inx}\,T(x) +
\frac{Q^2}{4}\,\,\delta_{n,0}~,~~~~~~~~~~~Q= b+\frac{1}{b}~, 
\end{equation}
and the central charge is 
$c=1+6Q^2$.
Using the free-field modes, from (\ref{T-phi^}) we find
\begin{eqnarray}\label{L0,pm}
&& 2b^2 \,L_0\,=\,\frac{p^2+(1+b^2)^2}{2}\,+\,a_{_{-1}}\,a_{_1}\,+\,
a_{_{-2}}\,a_{_2}\,+...~,
\\ \nonumber
&& 2b^2\, L_{\,1}=\left(p+i(1+b^2)\right)\,\,a_{\,_{1}}\,+
a_{_{-1}}\,a_{_2}+ a_{_{-2}}\,a_{_3}+...~,\\ \nonumber
&& 2b^2 \,L_{-1}=\left(p-i(1+b^2)\right)\,a_{_{-1}}\,+
a_{_{-2}}\,a_{_1}+ a_{_{-3}}\,a_{_2}+...~,
\end{eqnarray}
and similarly for other $L_n$'s. 
The $p$-dependent vacuum states $|p,\,0\rangle$ defined by (\ref{p-vacuum})
are obviously annihilated by $L_n$ ($n>0$) 
as well. The $SL(2,\rr)$ invariant vacuum is a special case, 
which is also annihilated by $L_0$ and $L_{-1}$. According to (\ref{L0,pm})
this state corresponds to $p=i+ib^2$. As we will see below the corresponding
state $|i(1+b^2),\,0\rangle$ usually is not normalizable, like the state
(\ref{|lambda>}) in the Morse potential for $\lambda>-\frac{\hbar}{2}$. 
The $SL(2,\rr)$ invariant state becomes
a bound state only for the special values of the boundary parameters
$l_b=r_b=-\cos\pi b^2$. In this case due to the symmetry 
$p\rightarrow -p$, one could also take the state with $p=-i(1+b^2)$. 
 However,   the state     
 $|-i(1+b^2),\,0\rangle$ is not annihilated by $L_{-1}$, if $a_{-1}$
is    treated as a standard creation operator in the Fock space. The 
key point for understanding of this subtlety is 
that $\phi(x)$ is not any more a free-field,
if $p$ becomes imaginary; though the Fourier mode expansion (\ref{phi-modes})
and the commutator relations (\ref{ccr}) are still valid. 
This issue was mentioned above
and it will be discussed in Section 5  in more
detail.

\subsection{The discrete spectrum of BLT}

Now we investigate the discrete spectrum  of the system  by the same scheme 
as in sub-section 3.1. 
The continuation of (\ref{<V>=}) to the sector of bound states 
is obtained similarly to (\ref{^V-b})
\begin{equation}\label{<V>-b}
\hat V= (2\sin \sigma)^{-\frac{b^2}{2}} \left (f(\theta)\,e^{-i(\theta-b^2)\tau}\,U_+
+U_-\,e^{i(\theta-b^2)\tau}\,
f(\theta)+g(\theta,\sigma)\right )~,
\end{equation}
where $f(\theta)=\sqrt{F(i\theta-ib^2)}$, 
$~g(\theta,\sigma)=G(i\theta,\sigma)$
and $U_\pm$ are the  raising-lowering  operators.
We choose again negative $\theta$. 
To get a $\theta $-spectrum bounded from below, candidates for ground states
$\theta _0$ must fulfill the equation $f(\theta _0)=0$. 
Furthermore, unitarity of the 
related Verma modules requires for the $\theta$-spectrum
\begin{equation}\label{red<theta<}
-(1+b^2)\leq \theta<0~,
\end{equation} 
where the lower bound corresponds to the $SL(2,\rr)$ invariant
situation ($1+b^2=bQ$). In the classical limit this interval corresponds to
$-1\leq \theta<0$, as discussed in section 2.1. 
Starting at such a ground state $\theta _0$ by acting with $\hat V$,
due to its raising part $U_+$, one reaches further states in integer steps
of $2b^2$. This generates a series with a finite number of discrete
$\theta$-values in the interval (\ref{red<theta<}). 
This series constitutes (a part of) the discrete spectrum iff there $\hat V$ is
Hermitean and positive. A necessary part of this condition is the
reality of $f$ and the positivity of $g$ at all points of the series except
the last one below the threshold at zero. In the following we analyze in some
detail this necessary part. At the end one then can check the positivity of all
eigenvalues of the matrix representing $\hat V$ on such a series, at least
numerically. 

To find our candidates for ground states we have to start with  the roots of
the equation $f(\theta)=0$.
From (\ref{F(p)=}) follows
\begin{equation}\label{f(theta)=}
f^2(\theta)=\frac{m_b^2\,\,\Gamma^{\,2}(1+b^2)}{\pi^2}\,\,
\Gamma(b^2-\theta)\Gamma(\theta-b^2)
\,\Gamma(-\theta)\,\Gamma(\theta-2b^2)
\,\,\Lambda(l_b,r_b;i\theta-ib^2)~,
\end{equation} 
and, therefore, the equation for the roots is reduced to
$\Lambda(l_b,r_b; i\theta-ib^2)=0$.
Then, by (\ref{u}) we find
\begin{equation}\label{^theta*}
\cos\pi(\theta_0-b^2)=-l_br_b\pm\sqrt{(1-l_b^2)(1-r_b^2)}~.
\end{equation}

In the classical case the lowest value of $\theta$ is obtained 
similarly from the equation $\Lambda(l,r;i\theta)=0$, 
which is quadratic with respect to
$\cos\pi\theta$, but one has to neglect the smaller root
and take only (\ref{theta_*}). The reason is that $\Lambda(l,r;i\theta)$ 
becomes negative in the interval between the roots, which leads to a complex
field in (\ref{V0}). From (\ref{theta_*})  $\theta_*$ is defined uniquely, 
because  classically it is bounded by $-1\leq\theta_*<0$.

In the quantum case there are, in general, more possibilities,
due to the discreteness of the spectrum and the deformations. To analyze these 
possibilities, it is convenient to introduce the following 
parameterization and notation 
\begin{equation}\label{lb-rb=}
l_b=\cos\pi\beta_l~,~~~~~~~r_b=\cos\pi\beta_r~;~~~~~~~~~
\beta_\pm=\beta_l\pm\beta_r~.
\end{equation}
To deal with the discrete spectrum we take the boundary parameters from the
interval $l_b,~r_b\in [-1,1]$. Then the parameters $\beta$ 
 are uniquely fixed if they
are chosen out of the basic intervals
\beq\label{basic-interval}
0\leq\beta_l,~\beta_r\leq 1~,~~~~~~~0\leq \beta _+\leq 2~,~~~~-1\leq\beta
_-\leq 1~.
\eeq 
From (\ref{u}) we get 
\begin{eqnarray}\label{f-cos}
\Lambda(l_b,r_b;i\theta-ib^2)&=&
4\cos \frac{\pi}{2}(\theta
-b^2+\beta_+)\cos \frac{\pi}{2}(\theta
-b^2-\beta_+) \\\nonumber
&&\times\cos\frac{\pi}{2}(\theta
-b^2+\beta_-) \cos\frac{\pi}{2}(\theta
-b^2-\beta_-)~.
\end{eqnarray}
Now the roots are given by
\begin{equation}\label{theta-0}
\theta_{0}~=~
1+b^2~-~\beta_{\pm}+2k~,~~~~~\mbox{or}~~~~~\theta_{0}~=~1+b^2~+~\beta_{\pm}+
2k~,
\end{equation}
with integer $k$.

The  candidate   higher levels are obtained by the action of the 
raising   operator $U_+$
\begin{equation}\label{theta-nm}
\theta ~=~1+b^2-\beta_{\pm}+
2k+2nb^2~,
~~~~~\mbox{or}~~~~~~~~\theta ~=~1+b^2+\beta_{\pm}+2k+2nb^2,~~~~~(n\geq 0)~.
\end{equation}
The condition (\ref{red<theta<}) imposes restrictions 
on the numbers ($k$, $n$). 
Other restrictions come from the Hermiticity
and positivity of $\hat V$, mentioned above.
The analysis of these conditions essentially depends on the value of $b^2$ 
and, in general, it is rather complicated. 
Note that for $b^2\geq 1$ only the case $n=0$ is allowed, and
for $b^2\leq 1$,  $k$  can not take more than one 
integer value, for each  of the four series in (\ref{theta-nm}). 

 We start with a discussion of the semi-classical situation $b^2\ll 1$. Then
 one can use the classical expressions
(\ref{F,G->hbar}) for $F$ and $G$,  what   simplifies the analysis.
In the classical theory  the lowest
allowed value for the boundary parameters is $-1$. 
In quantum theory the corresponding cases when $l_b$ and $r_b$ are near 
(in the units of $b^2$) to $-1$ need a separate
investigation,  which will be done below. For the moment let us  
 consider the case with  $b^2\ll 1-\beta_l$, $b^2\ll 1-\beta_r$ and $\beta_+>1$.
On the basis of the above described classical picture for the 
allowed lowest state  
we deduce that from the  spectrum candidates  
(\ref{theta-nm}) only the  series with $-\beta _+$ and $k=0$ remains and 
gives the spectrum 
\begin{equation}\label{^theta-n}
\theta_{n}=1+b^2-\beta_{+}+2nb^2~,~~~~(n\geq 0)~.
\end{equation}
Let us note   that the semi-classical calculation yields 
exactly  the same    answer \cite{DJ}.
\vspace{2mm}

 We now want to get at least some flavor of the setting beyond the 
semi-classical case. The analytic structure of $f^2$ in
(\ref{f(theta)=}) is governed by the poles of the $\Gamma $-factors and 
the zeros of $\Lambda(l_b,r_b;i\theta-ib^2)$ (see (\ref{f-cos})).
We restrict the following discussion to $b^2<1/3$. Then $f^2$ for generic
$\beta _-$ has only two
poles in the interesting $\theta$-interval (\ref{red<theta<}), namely at
$\theta = -1+b^2$ and $\theta =-1+2b^2$. In addition,
for $\beta _-=0$ the pole at $\theta = -1+b^2$ is canceled by a double
zero of  $\Lambda(l_b,r_b;i\theta-ib^2)$.  
\begin{figure}[h]
\begin{tabular}{cc}
\includegraphics[height=5.8cm]{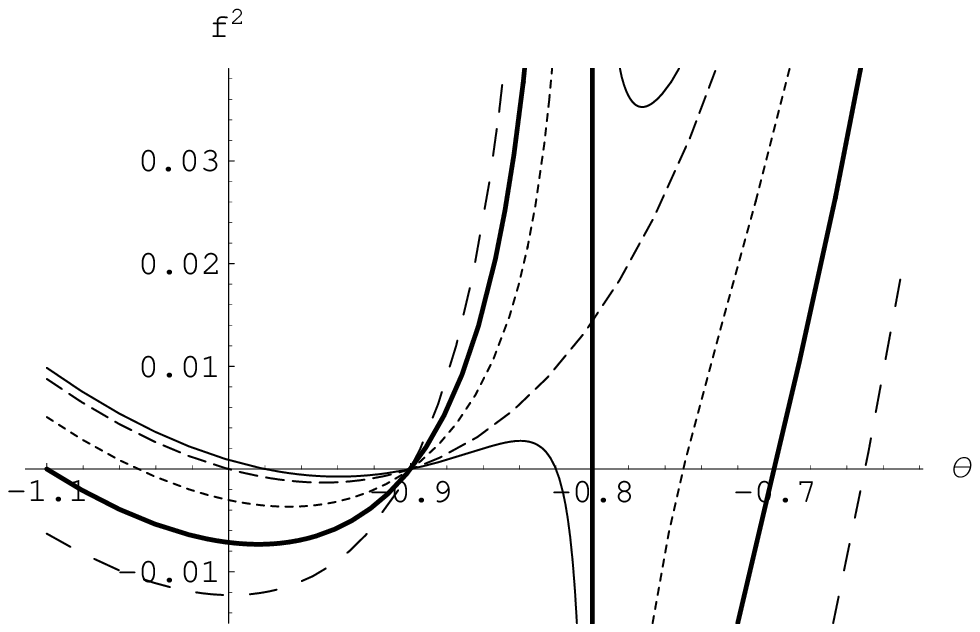} &
\includegraphics[height=4.8cm]{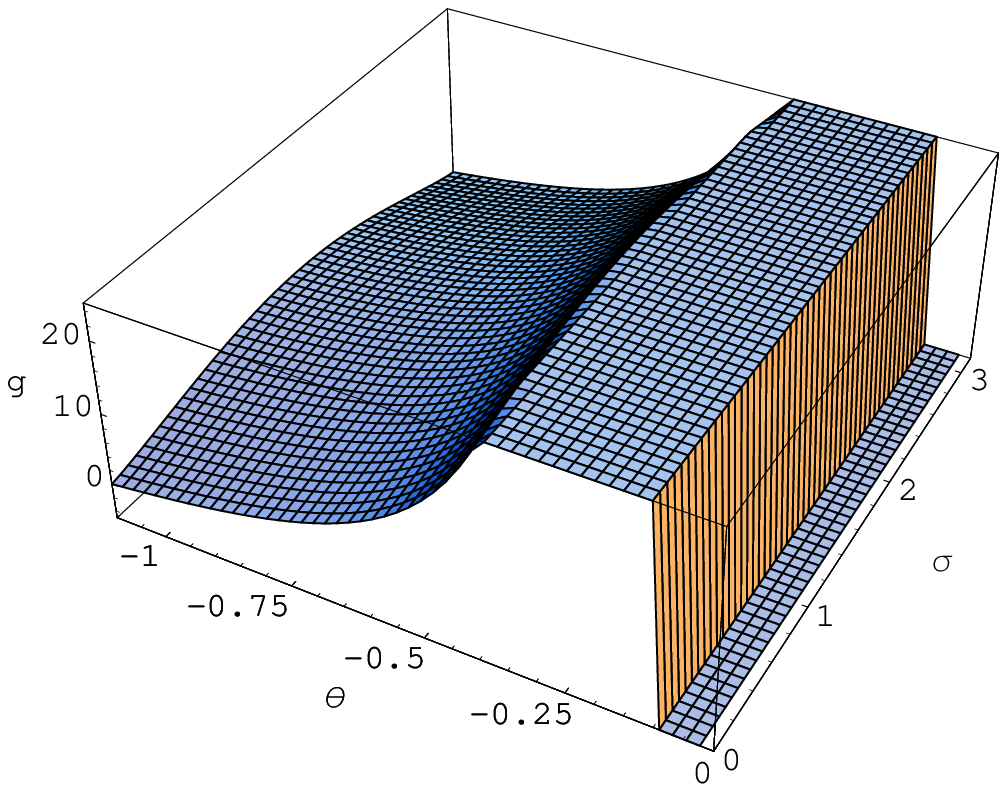} 
        \\[3mm](a) & (b)
\end{tabular}\\[5mm]
\noindent {\bf Fig.3}  (a) {\it The function $f^2(\theta )$ for
  $b^2=0.1,~\beta _-=0$\\
$~~~~~~~~~~~~~$ and
$\beta _+=1.75,~1.8~($i.e. $2-2b^2),~1.85,~ 1.9~($i.e. $2-b^2),~1.92$.}\\[1mm]
$~~~~~~~~~$ (b) {\it The function $g(\theta ,\sigma)$ for $b^2=0.1,~\beta
  _-=0,~ \beta _+=1.85$.}
\end{figure}

Let us consider first the symmetric case $l_b=r_b=\cos\pi\beta, ~\beta
_-=0,~\beta _+=2\beta $ with $0\leq
\beta\leq 1$. From a numerical analysis, for an example see fig. 3(b), 
we then conclude, that the function $g(\theta,\sigma)$ is positive for all
$\sigma $ and $-1-b^2<\theta<-b^2$. It has a pole at $\theta =-b^2$ and is
negative above this value. Since a point of the spectrum above $-2b^2$ would be
the last level below the continuum threshold, $\hat V$ has not to be 
well defined there, anyway. In conclusion, we get no obstruction for possible
spectral points from the properties of the function $g$ and can turn to the
analysis of $f^2$. For $\beta _-=0,~2b^2<\beta _+<2 $ from (\ref{theta-nm}) 
there remain only
three candidate series (for $0<\beta _+<2b^2 $ there is still another
option taking $k=-1$ in $-\beta _++2k$ in (\ref{theta-nm}))  
\bea\label{++theta-n,l=r}
\theta _n&=&1+b^2 -2\beta+2nb^2~,~\\
\label{--theta-n,l=r}
\theta _n&=&-1+b^2+2nb^2~,\\
\label{-+theta-n,l=r}
\theta _n&=&-3+b^2+2\beta + 2nb^2~.\,\,
\eea
These series can contribute to the spectrum only if $\theta _0$ is in the
interval (\ref{red<theta<}) {\it and} if $f^2\geq 0$ at $\theta _n,~n>1$.
\footnote{ For $f^2=0$ at $\theta _1$ the series would contain only $\theta
  _0$.\\ $~~~~~~~$A single point series would arise also for $-2b^2\leq \theta _0<0$.}
This implies
\bea\label{spec}
b^2<\beta\leq \frac{1+b^2}{2}~~:&&~~~~~\mbox{no discrete spectrum,}\\
\frac{1+b^2}{2}<\beta <1-b^2~~:&&~~~~~\mbox{spectrum given by
  (\ref{++theta-n,l=r}) only,}\nonumber \\
1-b^2\leq \beta <1 ~~:&&~~~~~\mbox{all 3 series
  (\ref{++theta-n,l=r})-(\ref{-+theta-n,l=r}) can be in the spectrum.}\nonumber
\eea
The described pattern is illustrated in fig.3(a) for $b^2=0.1$. 
As soon as $\beta _+>2-2b^2=1.8$ the function $f^2$ has three zeros, and
starting at each zero, after a jump $\theta\rightarrow \theta +2b^2=\theta
+0.2$ one reaches positive $f^2$.

An important special situation one finds for $\beta =1-b^2$. Then $\theta
_0=-1-b^2=-bQ$, corresponding to a $SL(2,\rr)$ invariant state, plays the role
of the ground state in the series (\ref{-+theta-n,l=r}). However, this state
is invariant under the action of $\hat V$, since $f^2$ is again zero
at $\theta _0+2b^2=-1+b^2$. This point just corresponds to the $\theta
_0$-value for the series (\ref{--theta-n,l=r}). Also the state at $-1+b^2$
is invariant, because after a second $2b^2$-jump we reach the zero of $f^2$
responsible for the ground state of the series (\ref{++theta-n,l=r}).
Therefore, at  $\beta =1-b^2$ we have several options: to take only the
spectrum given by (\ref{++theta-n,l=r}) or to take only one or both of 
the invariant states at $-1-b^2$ and $-1+b^2$ or to take (\ref{++theta-n,l=r})
plus one or both of the invariant states. 
Note that at $\beta=1$ all three series 
(\ref{++theta-n,l=r})-(\ref{-+theta-n,l=r}) coincide.

Closing the discussion of the $\beta_-=0$, case let us mention, that
also $\beta =1-b^2/2$ plays a special role. There $f^2$ has only 2 zeros and
the pole at $\theta =-1+2b^2$ is absent.
\vspace{2mm}
\begin{figure}[h!]
\begin{tabular}{cc}
\includegraphics[height=5.8cm]{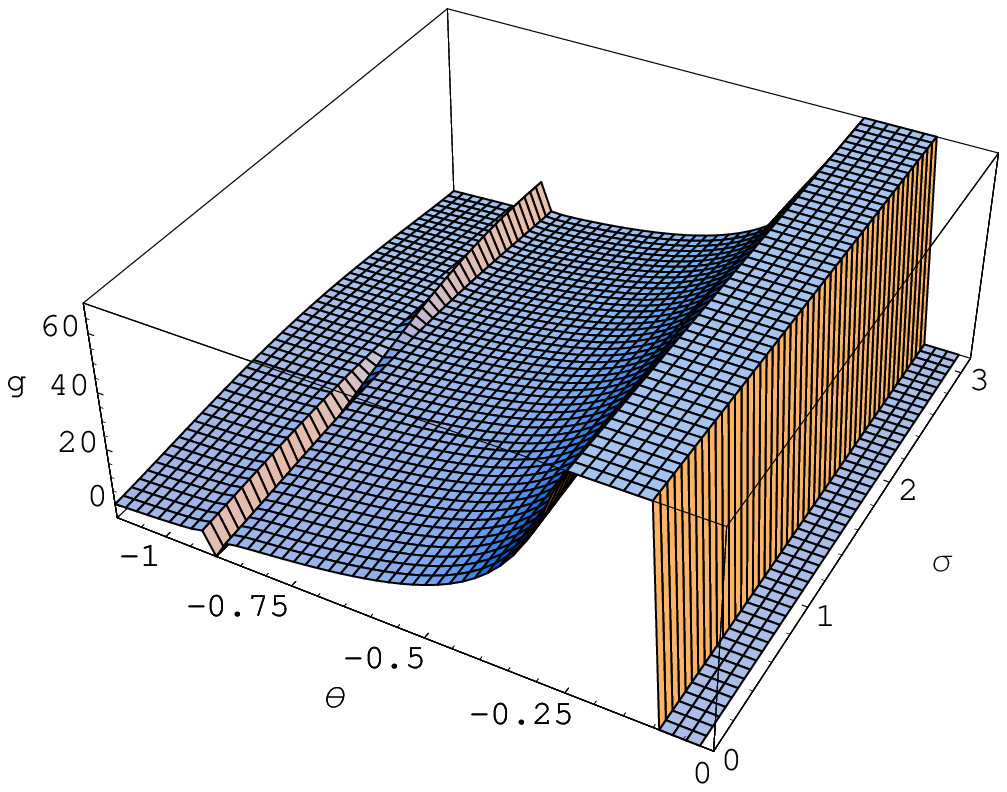} &
\includegraphics[height=5.8cm]{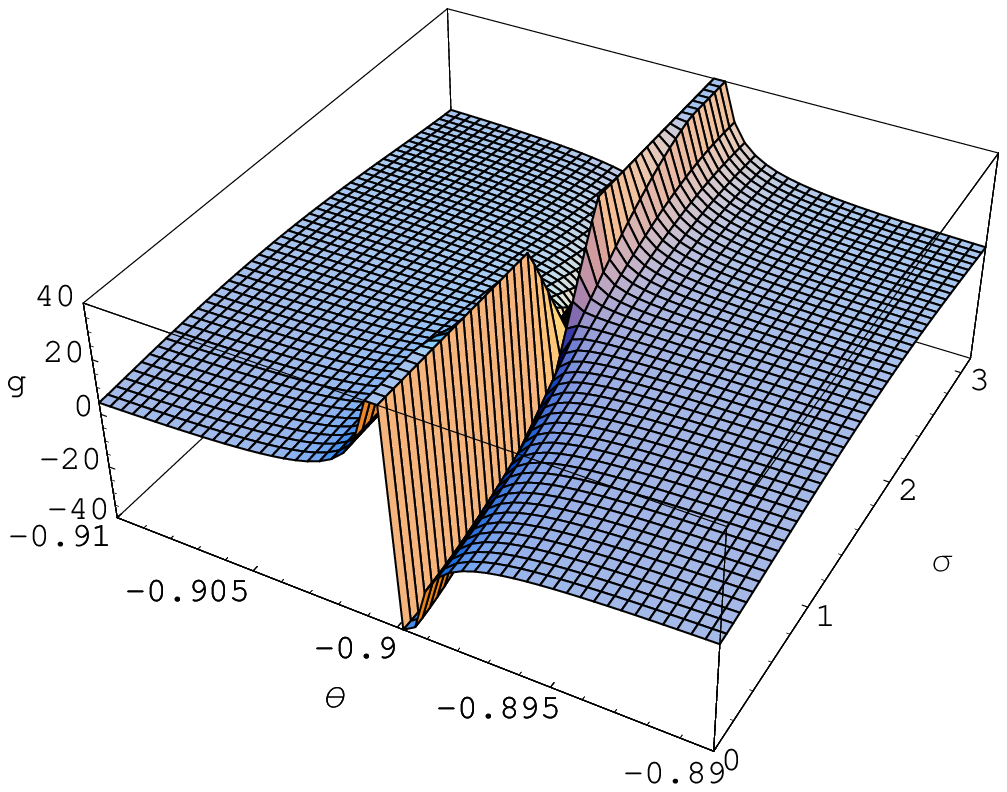}
\\[3mm](a) & (b)
\end{tabular}\\[5mm]
\noindent {\bf Fig.4} (a) {\it The function $g(\theta,\sigma)$ for
  $b^2=0.1,~\beta 
  _-=0.03,~\beta _+=1.92.$ Due to insufficient\\  
$~~~~~~~~~~~~~$resolution the plot shows the singularity at $\theta=-0.9$ in
  a rudimentary form only.  \\[1mm]
$~~~~~~~~~$(b) The same function, zoom into the singularity at $\theta=
  -1+b^2=-0.9$ .}
\end{figure}

The situation becomes more involved, as soon as  we switch on some asymmetry by
choosing $\beta _-\neq 0$. Then the function $g$ is no longer positive
definite in the whole range $-1-b^2<\theta<-b^2$, $0<\sigma<\pi$ and develops
a singularity at $\theta =-1+b^2$, for an example see fig.4.
The analysis has now to treat $f^2$ and $g$ in parallel and to take into
account the fact that $g$ can change sign as a function of $\sigma$. By
numerical study of several examples we found the remarkable possibility
that for $\beta_+ >2-2b^2$ one can reach situations where all 4 series
in (\ref{theta-nm}) can contribute to the spectrum. An example of this kind is
shown in fig.5. Starting at any of the four zeros of $f^2$ jumping a distance
$2b^2$ to the right one lands at positions with positive $f^2$ {\it and} $g$.
\begin{figure}[h!]
\begin{tabular}{cc}
\includegraphics[height=4.8cm]{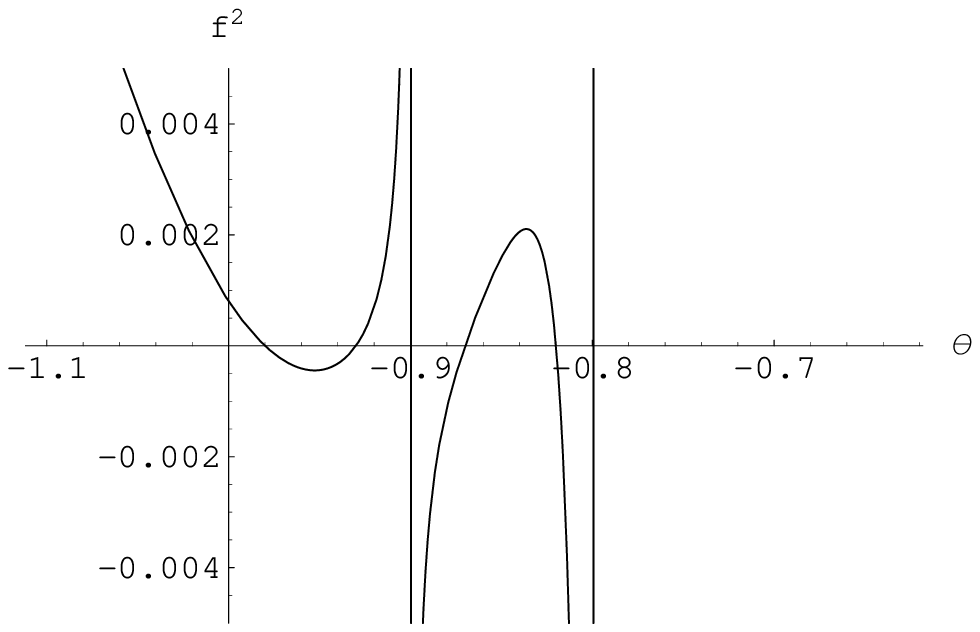}&
\includegraphics[height=4.8cm]{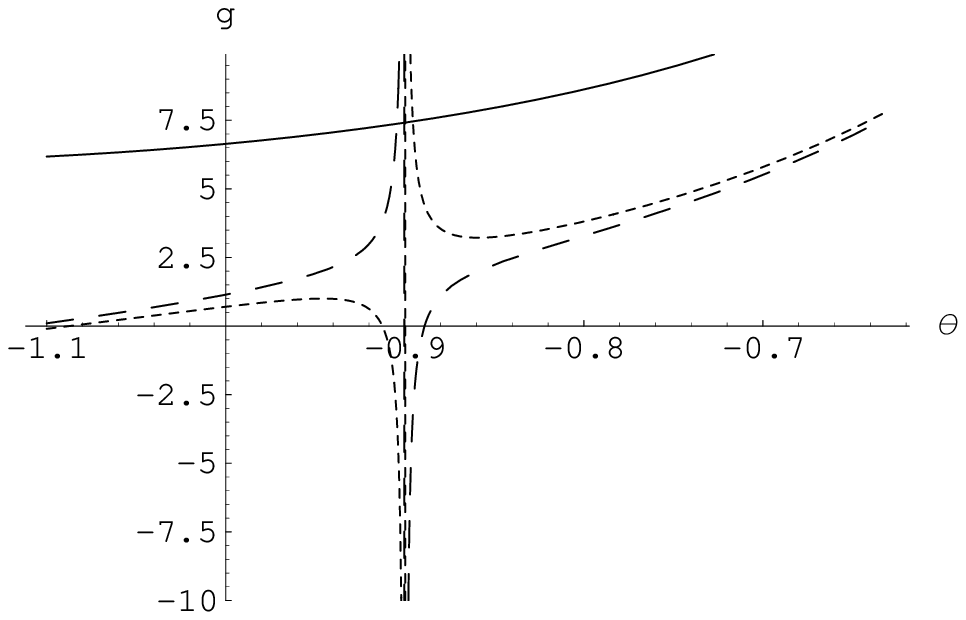}\\[3mm]
(a) & (b)
\end{tabular}\\[5mm]
\noindent {\bf Fig.5} (a) {\it The function $f^2(\theta)$ for $b^2=0.1,~\beta
  _-=0.03,~\beta_+=1.92$. For $\theta > -0.8,$ $f^2$ is\\
$~~~~~~~~~~~~~$positive.}\\[1mm]
$~~~~~~~~~~$(b) {\it The function $g(\theta )$ for $b^2=0.1,~\beta
  _-=0.03,~\beta _+=1.92,$\\$~~~~~~~~~~~~~$ and $\sigma = 0$ (dashed), $ \pi
 /2$ (full), $\pi $ (dotted),
 respectively.}
\end{figure}
At the end of these studies we mention, that for all examples
plotted in the figures the positivity of all eigenvalues of $\hat V$ has been
checked numerically.

When there are different options for the spectrum it
is natural to assume that all of them are realized. This conjecture fixes the
spectrum and, respectively, the $\hat V$-operator in the bound state sector. 
Let us note again that different options arise if
values of the boundary parameters are near to the critical ones. 
In the next section we investigate how the different possibilities for the
spectrum are realized in terms of the reflection amplitude.   

\subsection{The reflection amplitude of BLT}

Similarly to  (\ref{R-eq}),  the reflection amplitude $R(p)$
associated with the operator 
(\ref{<V>=}) satisfies the  equation 
\begin{equation}\label{R(p)}
R\left(p+ib^2\right)= R\left(p-ib^2\right) F(p)~,
\end{equation}
where $F(p)$ is given by (\ref{F(p)=}). 
To find $R(p)$ we apply the scheme 
described in the sub-section 3.2, which provides an 
integral representation for the phase $\gamma(p)=2ib^2\log R(p)$.
At the first stage we have to find a Fourier type integral representation
for $\log F(p)$. The $\Gamma$-factors standing in $F(p)$ can be represented
with the help of (\ref{log-Gamma=}), and for
$\Lambda(l_b,r_b;p)$ we use the continuation of (\ref{f-cos}) to the 
scattering sector
\begin{equation}\label{up=}
\Lambda(l_b,r_b;p)=4\,\cos\frac{_\pi}{^2}(ip-\beta_{_+})\,
\cos\frac{_\pi}{^2}(ip+\beta_{_+})\, \cos\frac{_\pi}{^2}(ip-\beta_{_-})\,
\cos\frac{_\pi}{^2}(ip+\beta_{_-})~.
\end{equation}
Here $\beta_\pm$ are the angle variables (\ref{lb-rb=}), which
for $ l_b>1$ and $r_b > 1$ become pure imaginary. Writing then the $\cos$-terms 
of (\ref{up=}) through $\Gamma$-functions by (\ref{Gamma-Gamma=ch}), 
we can apply  again (\ref{log-Gamma=}).
Other technical details are given in Appendix C, and we obtain 
\begin{eqnarray}\label{gamma(p)=1}
&&\gamma(p)=p\log\left(4\pi^2m_b^2\Gamma^2(1+b^2)\right)+\\ \nonumber
&&\int_0^\infty \frac{dt}{t}\left[\frac{2ib^2\left(e^{-ipt}-e^{ipt}\right)}
{(1-e^{-b^2t})\left(1-e^{-t}\right)}
-{p}e^{-t}\left({4+2b^2}+\frac{4}{1-e^{-t}}\right)\right]+\\ \nonumber
&&\sum_{\varepsilon,\,\nu}\int_0^\infty \frac{dt}{t}\left
[\frac{2ib^2\left(e^{ipt}-e^{-ipt}\right)e^{-(1+b^2-
\varepsilon\beta_\nu)t}}
{\left(1-e^{-{2b^2 t}}\right)\left(1-e^{-2t}\right)}+p\left(
\frac{2e^{-2t}}{1-e^{-2t}}+(1+\varepsilon\beta_\nu)
e^{-2t}\right)\right]~.~~~~~
\end{eqnarray}
To represent these integrals in a familiar form, we rescale the integration 
variable
for the term in the first line 
by  $t\mapsto \frac{t}{b}$ and for the  other four terms (which are 
given as a sum over $\varepsilon=\pm 1$ and $\nu\equiv\pm$) 
by $t\mapsto \frac{t}{2b}$. Then,
using (\ref{Gamma-b}) and the integral (\ref{log(2pi)}), 
we express (\ref{gamma(p)=1}) 
through the Double Gamma function and the reflection amplitude becomes
\begin{eqnarray}\label{R(p)=}
R(p)={\cal R}(s_{_+},s_{_-}; P)\equiv
\left[m_b^2\,\Gamma^2(b^2)\,b^{2-2b^2}\right]^{-\frac{iP}{b}}\,
\frac{\Gamma_b\,(2i\,P\,)}{\Gamma_b(-2iP)}\times
~~~~~~~~~~~~~~~~~~~~~~~~~~
\\ \nonumber
\frac{\Gamma_b(\frac{Q}{2}-iP+s_{_+})\Gamma_b(\frac{Q}{2}-iP-s_{_+})
\Gamma_b(\frac{Q}{2}-iP+s_{_-})\Gamma_b(\frac{Q}{2}-iP-s_{_-})}
{\Gamma_b(\frac{Q}{2}+iP+s_{_+})\,\,\Gamma_b(\frac{Q}{2}+iP-s_{_+})\,
\Gamma_b(\frac{Q}{2}+iP+s_{_-})\,\Gamma_b(\frac{Q}{2}+iP-s_{_-})}~,
\end{eqnarray}
where $Q$ is the background charge (\ref{l_n}) and the variables $P$ and
$s_{_\pm}$ are obtained by the rescalings  
\begin{equation}\label{->zz1}
P=\frac{p}{2b}~,~~~~~~~~~~~~~~s_{_\pm}=\frac{\beta_{_\pm}}{2b}~.
\end{equation}
In this way we come to the notations of \cite{FZZ} and to 
have a complete correspondence, 
we also relate the Liouville fields and mass and boundary 
parameters,   
which are fixed by the normalization of the action (\ref{S})
\begin{equation}\label{->zz}
b \phi    =\varphi~,~~~~~\pi b^2\,\mu= m^2~,~~~~~
2\pi b^2 \mu_{_{B_1}}=2ml~,~~~~~2\pi b^2 \mu_{_{B_2}}=2mr~.
\end{equation}
Here the left hand sides denote the corresponding quantities of
\cite{FZZ}.  
In  this  paper we prefer to work with the initial 
action (\ref{S})  without the rescalings (\ref{->zz}). 
It helps us to control the semi-classical behavior of the theory
and guides  to fix results of quantum calculations.

To match eq. (2.25) of \cite{FZZ}: $\cosh\pi b s_1=\mu_{\scriptsize B_1}\mu ^{-1/2}\,\sqrt{\sin\pi b^2}$
(and similarly for $s_2$), we introduce $s_1$ and $s_2$ by:
$s_\pm=\frac{i}{2}(s_1\pm s_2)$ and use (\ref{lb-rb=}) together with
(\ref{->zz1})-(\ref{->zz}) and (\ref{m-m_b}).
As a result, $R(p)$ given by (\ref{R(p)=}) 
coincides with
the reflection amplitude of BLT discussed in \cite{FZZ, T,BRZ}.

However, the integral representation 
(\ref{gamma(p)=1}), as it is derived in Appendix C, 
assumes $b^2< 1$ and $1-\varepsilon\beta_\nu>0$
for all four pairs ($\varepsilon, \nu$).  The integrals in (\ref {gamma(p)=1})
are also well defined for pure imaginary $\beta_\pm\,$ $\,\,(l_b>1$,
$r_b>1)$. Both  cases have no bound states and, therefore, the reflection
amplitude (\ref{R(p)=}) for them is exact.
Then, similarly to the Morse potential,
the reflection amplitude of BLT with discrete spectrum 
is obtained by a continuation of  (\ref{R(p)=}) in the boundary parameters, 
until they reach the critical value.
\vspace{2mm}

Let us  discuss the continuation issue 
for the symmetric boundary conditions $l_b=r_b$. 
For $l_b\geq 1$, with the parameterization $l_b=\cosh\pi\rho$ ($\rho \geq 0$), 
eq. (\ref{R(p)=}) becomes 
\begin{equation}\label{R_p,l>1}
 R(p)={\cal R}(is,0; P)~,
\end{equation}
where $s=\frac{\rho}{b}$, $P$ is given by (\ref{->zz1}) and the function 
${\cal R}(is,0; P)$ reads
\begin{eqnarray}\label{R(p),l=r}
{\cal R}(is,0; P)&=&\left[m_b^2\Gamma^2(b^2)b^{2-2b^2}\right]^{-\frac{iP}{b}}
\times
\\ \nonumber
&&\frac{\Gamma_b\,(2i\,P)}{\Gamma_b(-2iP)}\,
\frac{\Gamma_b(\frac{Q}{2}-iP+is)\Gamma_b(\frac{Q}{2}-iP-is)
\Gamma_b^2(\frac{Q}{2}-iP)}
{\Gamma_b(\frac{Q}{2}+iP+is)\Gamma_b(\frac{Q}{2}+iP-is)
\Gamma_b^2(\frac{Q}{2}+iP)}~.
\end{eqnarray}
In the next interval $0\leq l_b< 1$, using the standard parameterization
$l_b=\cos\pi\beta$  with $\,\,0<\beta\leq \frac{1}{2}$, we have again
(\ref{R_p,l>1}), but now with $is=\frac{\beta}{b}$.
The $s$-dependent part in (\ref{R(p),l=r}) is given by  
\begin{equation}\label{X(s)}
\frac{\Gamma_b(\frac{Q}{2}-iP+is)\Gamma_b(\frac{Q}{2}-iP-is)}
{\Gamma_b(\frac{Q}{2}+iP+is)\Gamma_b(\frac{Q}{2}+iP-is)}=
\exp\left[i\int_0^\infty \frac{dt}{t}\left(\frac{\sin(2Pt)\,\cos(2st)}
{\sinh(bt)\,\sinh(t/b)}-\frac{2P}{t}\right)\right]~.
\end{equation}
Note that the parameter $s$ does not change smoothly when we pass the point
$l_b=1$, but the  dependence of (\ref{X(s)}) on $l_b$ is analytic.
Farther continuation to the interval $-1\leq l_b<0$ corresponds to 
$\frac{1}{2}<\beta\leq 1$.
The integral in (\ref{X(s)}) becomes divergent
for pure imaginary $s$, if $|is|>\frac{Q}{2}\,$ 
$\,\,(\beta>\frac{1+b^2}{2}).\,$ Then one has to use the 
analytical continuation of (\ref{R(p),l=r}) provided 
by (\ref{Gamma_b(z+1/b)}). 

To get the discrete spectrum from the reflection amplitude one has to 
look for its zeros on the imaginary
axis $p=i\theta$, with  bounded $\theta$:  $-1-b^2\leq\theta<0$. 
The function $\Gamma_b(z)$ has no zeros, like usual 
$\Gamma$-function, but it has
poles at $z=-\frac{m}{b}-nb\,$ (\,$m\geq 0$, $n\geq 0$).
One has to note that not all zeros of $R(i\theta)$
correspond to the discrete spectrum. For example,
the reflection amplitude for the Morse potential
(\ref{R^s}) has zeros at $p=-\frac{in\hbar}{2}$ ($n\,\geq 0$), 
corresponding to the poles of $\Gamma\left(-\frac{2ip}{\hbar}\right)$, but
they are not in the spectrum, and the spectrum 
is obtained by the poles of another factor in the denominator, which is 
$\lambda$ dependent. 
Similarly, in (\ref{R(p),l=r})  one has to take only
$s$-dependent zeros, corresponding to the poles of $\Gamma_b(\frac{Q}{2}+
iP-is)$. They define exactly the spectrum 
(\ref{++theta-n,l=r}).

As it was mentioned in the previous subsection, there are other
possibilities for the discrete spectrum near the critical value of
the boundary parameter $\beta\geq 1-b^2$.
Some candidates for the corresponding reflection amplitude  can be represented
in the form (\ref{R(p)=}), with a suitable choice of the pair 
$(s_{_+},\,\,s_{_-})$.
For example, the case $(s_{_+}=\frac{\beta}{b}$, $s_{_-}=\frac{1}{b})$
corresponds to the unification of the spectra (\ref{++theta-n,l=r}) 
and (\ref{--theta-n,l=r}). To unify (\ref{--theta-n,l=r}) 
and (\ref{-+theta-n,l=r}),
one has to take $(s_{_+}=\frac{2}{b}-\frac{\beta}{b}$, $s_{_-}=\frac{1}{b})$;
and the case $(s_{_+}=\frac{2}{b}-\frac{\beta}{b}$, $s_{_-}=0)$ describes
only the spectrum (\ref{-+theta-n,l=r}). But these examples do not cover
the case with all three spectra (\ref{++theta-n,l=r})-(\ref{-+theta-n,l=r}), 
since (\ref{R(p)=}) can not give more than two series of equidistant spectra.
A modified reflection amplitude, which covers all three series 
(\ref{++theta-n,l=r})-(\ref{-+theta-n,l=r}),
is given by 
 \begin{equation}\label{R-3spectra}
R(p)={\cal R}\left(\frac{\beta}{b},\frac{1}{b}; P\right)\,\,\,
\frac{\sin\frac{\pi}{b}
(\frac{Q}{2}+iP-\frac{2-\beta}{b})}
{\sin\frac{\pi}{b}
(\frac{Q}{2}-iP-\frac{2-\beta}{b})}~.
\end{equation}
It has unit norm and satisfies eq. (\ref{R(p)}), since the ratio 
of $\sin$-functions 
is invariant under the shift $P\mapsto P-ib$. Such a ratio of
$\sin$-functions  modifies the reflection amplitude (\ref{R(p)=}) in a form
compatible
with the more general spectrum discussed in the previous sub-section. 

For $\beta=1-b^2$ the reflection amplitude (\ref{R-3spectra}) becomes 
 \begin{equation}\label{R-zz}
R(p)={\cal R}\left(\frac{1}{b}-b,\frac{1}{b}; P\right)\,\,\,
\frac{\sin\frac{\pi}{b}
(\frac{Q}{2}-iP)}
{\sin\frac{\pi}{b}(\frac{Q}{2}+iP)}=
{\cal R}\left(\frac{1}{b}-b,0; P\right)\,\,\,\frac{\sin^2\frac{\pi}{b}
(\frac{Q}{2}-iP)}
{\sin^2\frac{\pi}{b}(\frac{Q}{2}+iP)}~,
\end{equation}
Here, using the property (\ref{Gamma_b(z+1/b)}) 
of $\Gamma_b$-functions, we have shifted the $s_{_{-}}$ argument
of the function ${\cal R}$ from $\frac{1}{b}$ to $0$. This creates
an additional ratio of $\sin$-functions. 
Note that the reflection amplitude (\ref{R-zz}) indeed vanishes for
$P=-\frac{iQ}{2}$,
which corresponds to the  case with a normalizable $SL(2.\rr)$ invariant 
vacuum.

\subsection{1-point function of BLT}

The $1$-point function of BLT associated
with the  operator $V$ is given by
the matrix element of $V$ between the
$SL(2,R)$ invariant vacuum states.
   The state  $|ibQ, 0\rangle$, as    a  continuation of the states 
$|p, 0\rangle$ to     $p=i(1+b^2)=ibQ$,  
is $SL(2,R)$ invariant, but for generic $l_b,~r_b$ defined only as    a 
singular element of
the dual (to $L^2(\rr_{_+})$) space, like the state (\ref{|lambda>}) 
for $\lambda>-\frac{\hbar}{2}$.     Therefore, one has to be careful
in giving a well defined meaning to the 1-point function.

\subsubsection{ZZ case, $l_b=r_b=-\cos\pi b^2$} 
In this case  the state $|ibQ, 0\rangle$ is normalizable since,
according to subsection 4.1, $-bQ$ is in the $\theta$-spectrum. 
Let us call the corresponding normalized state by
$\vert {\cal O}\rangle $. Then we get with (\ref{<V>-b}) and (\ref{G(p)=})
\begin{equation}\label{<theta_0|V|theta_0>}
\langle {\cal O}|V|{\cal O}\rangle     =(2\sin\sigma)^{-\frac{b^2}{2}}\,\,
G(p,\sigma)|_{p=i+ib^2}~.
\end{equation}
    
For the evaluation of $G$ we note that the coefficient $b_p$ vanishes 
for $l_b=r_b$ and  $p=i+ib^2$, and $c_p$ becomes (see (\ref{bp}))
\begin{equation}\label{b_p=}
c_p=-\frac{2i\,m_b\,l_b\,e^{i\pi b^2}}{\pi}\,\,
\Gamma^2(1+b^2)\Gamma(-1-2b^2)\,\,\sin\pi b^2~.
\end{equation}
At the same time, the hypergeometric function standing  in (\ref{<Vc>=})
reduces to 
 \begin{equation}\label{p=-i}
_2F_1\left(-b^2,
-1-2b^2,-b^2;e^{2i\sigma}\right)=\left(1-e^{2i\sigma}\right)
^{1+2b^2}~.
\end{equation}
As a result we obtain
\begin{equation}\label{p=-i,<V>}
    \langle {\cal O}|V|{\cal O}\rangle = - 2m_b\, \cos\pi b^2 
   ~
\frac{\Gamma(1+b^2) \,\Gamma(-1-2b^2)}
{\Gamma(-b^2)}\,\left(2\sin\sigma\right)
^{1+\frac{3}{2}\,b^2}~.
\end{equation}

This result agrees with the 1-point function of \cite{ZZ} for the special case
corresponding to our $V$ and the 'basic' (1,1) vacuum. To make this
manifest, one has to relate $m_b$ via (\ref{->zz}) and
(\ref{m-m_b}) to their mass parameter and to take into account also the
relative rescaling of the Liouville field. 

\subsubsection{FZZT case, generic $l_b=r_b$} 

In this case the $SL(2,\rr)$ invariant vacuum is not normalizable and
to calculate the 1-point correlation function corresponding to the operator
(\ref{<V>=})  we need a regularization procedure like the one given
at the end of sub-section 3.3. This procedure for BLT needs further
investigation and will be discussed elsewhere.

Assuming for the time being that eq. (\ref{R'}), derived for the Morse
 potential, is ana\-lo\-gously valid also in BLT for the operator 
 (\ref{<V>=}), we get 
\begin{equation}\label{R'1}
_{\nn}\langle ibQ,0|V|ibQ,0\rangle_{\nn}=
(2\sin\sigma)^{-\frac{b^2}{2}} G(ibQ,\sigma)\,\,
\left[-2ib^2\,\partial_pR(p)\right]_{p=ibQ}~.
\end{equation}
The factor $(2\sin\sigma)^{-\frac{b^2}{2}} G(ibQ,\sigma)$ is obtained from 
the r.h.s. of (\ref{p=-i,<V>}), after replacing the term $-\cos\pi b^2$ 
by the generic boundary parameter $l_b=\cosh\pi bs$. 
Switching to the parameters $m,l$
(see  (\ref{m-m_b})), this gives  
\beq\label{1pointFZZ}
(2\sin\sigma)^{-\frac{b^2}{2}} G(ibQ,\sigma) 
~=~Z(b)~(2\sin\sigma )^{1+\frac{3}{2}b^2}~,
\eeq
with $Z(b)$ defined in (\ref{bulk-bound-str}) as a bulk-boundary structure
constant.

Using the reflection amplitude (\ref{R(p),l=r}), from (\ref{R'1}) we find
\begin{eqnarray}\label{R'2}
\frac{_{\nn}\langle
 ibQ,0|V|ibQ,0\rangle_{\nn}}{(2\sin\sigma)^{1+\frac{3}{2}\,b^2}}
=\left(m_b^2\,\Gamma^2(b^2)\right)^{1+\frac{1}{2b^2}}\,
\Gamma(-2b^2)\,
\Gamma\left(-2-\frac{1}{2b^2}\right)\times\nonumber \\
\left[\cosh\pi s\left(2b+\frac{1}{b}\right)-
\cosh\pi s\left(2b-\frac{1}{b}\right)\right]~.
\end{eqnarray}
The r.h.s. of this equation can be compared with the 1-point function
$U(\alpha)$ of \cite{FZZ}
\beq \label{<V>FZZ}
U(\alpha
)=\frac{2}{b}
\left(\frac{\pi\mu\,\Gamma(b^2)}{\Gamma(1-b^2)}\right)^{\frac{Q-2\alpha}{2b}}
\Gamma\left(2b\alpha-b^2\right)\,\Gamma\left(2\alpha /b-1/b^2 -1
\right)\,\cosh\pi s(2\alpha -Q)~.
\eeq 
at $\alpha=-\frac{b}{2}$. The translations between the parameters (\ref{->zz})
and (\ref{m-m_b}) provide
\beq \label{alpha=-b/2}
U\left(-\,{b}/{2}\right)=\frac{2}{b}
\left(m_b^2\,\Gamma^2(b^2)\right)^{1+\frac{1}{2b^2}}\,
\Gamma(-2b^2)\,
\Gamma\left(-2-\frac{1}{2b^2}\right)\,
\cosh\pi s\left(2b+\frac{1}{b}\right)~.
\eeq 
The pre-factor $2/b$ in this equation could be related to the relative
normalization of the states $|P\rangle$ and $|p\rangle$,
 but the existence of  the second $\cosh$-term in (\ref{R'2})
indicates that the rule (\ref{R'2}) yields a 1-point function 
not compatible with the conformal bootstrap result (\ref{<V>FZZ}). 

Furthermore, replacing (\ref{R'2}) by the weaker assumption of some
factorization 
\begin{equation}\label{R'3}
_{\nn}\langle ibQ,0|V|ibQ,0\rangle_{\nn}=N\,\,
(2\sin\sigma)^{-\frac{b^2}{2}} G(ibQ,\sigma)~,
\end{equation}
where $N$ denotes a regularized norm $_{\nn}\langle ibQ,0|ibQ,0\rangle_{\nn}$,
we can identify $N$ with $U(0)$, and then, due to (\ref{1pointFZZ}),
$Z(b)$ should be compared with 
\beq\label{ratio}
\frac{U(-b/2)}{U(0)}~=~m\sqrt{\frac{\sin\pi b^2}{\pi b^2}}~\frac{\cosh (b+Q)\pi
    s}{\cosh Q\pi s}~\frac{\Gamma (1+b^2)\Gamma (-1-2b^2)}{\Gamma (-b^2)}~.  
\eeq
Obviously, this does not agree with our $Z(b)$. However, there is an
intriguing observation concerning the construction of $U(\alpha )$ as a
solution of a functional equation in \cite{FZZ}. Replacing $Q$ in the argument
of $\cosh $ in (\ref{<V>FZZ}) by an arbitrary constant $c$ gives still a
solution of this  functional equation. Using the freedom to put $c=0$
one would get for $U(-b/2)/U(0)$ just $1/2\cdot Z(b)$. The choice $c=Q$ is
forced  by implementing the reflection relation as an additional input of the
bootstrap approach.  

Altogether, at the present status of our calculations,
it is either not possible to make a comparison, due to the breakdown of
(\ref{R'2}), or the factorization (\ref{R'3}), or 
there is a deeper problem in relating Euclidean bootstrap
BLT to Lorentzian operator BLT.

\section{Some open problems of the operator approach to BLT}

Concerning open problems of the operator approach, first we refer 
to the expansion of the vertex operators in powers of the
screening charges (\ref{V_alpha1}). The same expansion in the periodic case
contains only even powers of $m_b$ and the general $p$-dependent coefficient
is known in a closed form \cite{OW}. 
This enables one to verify that the operator
Liouville equation (\ref{^L-eq}) is fulfilled by the formal power series of
$\varphi(x,\bar x)$ and $V_1(x,\bar x)$. However, in general, these expansions  
are asymptotic and, therefore, ineffective in practical calculations. But for
$\alpha=-\frac{n}{2}$ the series becomes finite, as it is expected from the
classical picture. $V_{-\frac{n}{2}}$ contains only $n+1$ terms and one can
obtain the corresponding correlation function 
$\langle p',\,0|V_{-\frac{n}{2}}|p,\,0\rangle$ in a closed form \cite{JW}. The
continuation of this expression to arbitrary $\alpha$ reproduces
the $3$-point correlation function of \cite{DO, ZZ1}. 
We hope that this scheme of
calculation of the correlation functions could work effectively for BLT as
well, if one finds the coefficients $c_p^{l,k}(\alpha)$ in a closed form.
Note that the operators $V_{-\frac{n}{2}}$ for BLT
were constructed in \cite{GK}, using a quantum group structure of the basic
operators. But a representation of that result as a polynomial in powers of
screening charges needs an additional 
combinatorial labor similar to the one given
in Appendix A. The power series for the operators $V_\alpha$ can also be used
to find its compact integral representation \cite{T1,JW}, which is directly
related to the correlation functions.

Another issue we would like to discuss is the $S$-matrix of Liouville theory, 
which is unknown in a closed form  for the periodic case as well, though the 
expression for the $out$-field exponential in terms of the $in$-field is 
rather compact (\ref{^Vout}). This relation provides the following
equation for the $S$-matrix
\begin{equation}\label{^S}
e^{-\phi(\bar x)}\,\,e^{-\phi(x)}{\cal S}=m_b^2\,{\cal S}
\,A(\bar x)\,e^{-\phi(\bar x)}\,D_p\,e^{-\phi(x)}\,A(x)~.
\end{equation}
On the basis of the classical picture, the $S$-matrix can be represented in the
form ${\cal S}={\cal P}\,R(p)\,{S}_p$, 
where ${\cal P}$ is the parity operator in
the zero mode sector as in (\ref{S-eq}), $R(p)$ is the reflection amplitude
\begin{equation}\label{Rp}
{\cal S}\,|p,\,0\rangle=R(p)\,|-p,\,0\rangle~,
\end{equation}
and  ${S}_p$ is the operator responsible for transitions
in non-zero mode sectors. $S_p$ contains $p$ as a parameter
and depends on the $a_n$ operators. Inserting
this ansatz in eq. (\ref{^S}) and projecting it between the vacuum states, 
one finds that $R(p)$ indeed satisfies (\ref{R(p)}). 
Since $R(p)$ is known, eq. (\ref{^S}) reduces to
a closed equation for ${S}_p$
\begin{equation}\label{^S_p}
s(p)\,e^{-\hat\phi(x)}\,{S}_p=\,{S}_{p-ib^2}
\,\int_0^{2\pi}dy\, e^{p(y-\pi)}\, e^{2\hat\phi(x+y)}\,\,e^{-\hat\phi(x)}~,
\end{equation}
where the function $s(p)$ is the integrated short distance factor 
\begin{equation}\label{s_p}
s(p)=\int_0^{2\pi}dy\,e^{p(y-\pi)} (1-e^{-iy})^{b^2}=
\frac{2\pi\,\Gamma(1+b^2)}
{\Gamma(1- ip)\Gamma(1+b^2+ ip)}~,
\end{equation}
and $e^{-\hat\phi(x)}$ and $e^{2\hat\phi(x+y)}$ denote the normal ordered
chiral free-field exponentials without the zero modes
$e^{-\hat\phi(x)}=e^{-\phi_+(x)}e^{-\phi_-(x)}$. Note that in the periodic
case the $S$-matrix has a similar form ${\cal S}={\cal P}\,R(p)\,{S}_p\,\bar
S_p$, where the anti-chiral part  $\bar S_p$ (depending on $\bar a_n$)
is functionally identical to $S_p$, and $S_p$ satisfies the same equation
(\ref{^S_p}). From the analysis of this equation
one can conclude that the transition amplitudes defined by $S_p$ are non-zero
only between equal levels of the Hamiltonian $L_0$ and
$S_p$ has the following structure
\begin{eqnarray}\label{S_p=}
S_p=I+R^{1,1}_p\, a_{_{-1}}\, a_{_1}+
~~~~~~~~~~~~~~~~~~~~~~~~~~~~~~~~~~~~~~~~~~~~~~~~~~~~~~~~~~~~~~~~~~~~~~~~~~~\\ 
\nonumber
[R^{01,01}_p\, a_{_{-2}}\, a_{_2}+R^{01,20}_p\, a_{_{-2}}\, a^2_{_1}+
R^{20,01}_p\,{a^2_{_{-1}}}\, a_{_2} +
R^{20,20}(p)\,{a^2_{_{-1}}}\,{a}_{_1}^2]+\dots~,
\end{eqnarray}
with some $p$-dependent coefficients $R^{1,1}_p$, $\,R^{01,01}_p,\dots$.

Projecting eq. (\ref{^S_p}) 
between  the states of the Fock space, one finds the $p$-dependent
coefficients and transition amplitudes
step by step. For examples, the projection between the vacuum and the first
exited states provides
\begin{equation}\label{<1|S|1>}
S_p\,\,a_{_{-1}} |p, 0\rangle = - 
\frac{p+i(1+b^2)}{p-i(1+b^2)}\,\,a^+_{_{1}}|p, 0\rangle~.
\end{equation}
The aim is to find the matrix element of $S_p$ between arbitrary
coherent states, which provides the normal symbol, or normal ordered
form $S_p$.  
From eq. (\ref{^S_p}) one can find matrix elements between some 
coherent states, but a closed form of $S_p$ is still missing.

Finally we consider the problem of quantization of the sector
of bound states, which corresponds to pure imaginary $p$. The Liouville field
exponential (\ref{V-xi}) for $p=i\theta$ (negative $T_0$) can be written as
\begin{eqnarray}\label{V(x,bar x),p=itheta}
V(x,\bar x)=e^{-\phi(x)}\,e^{-\phi(\bar x)}+e^{-\phi^*(x)}\,e^{-\phi^*(\bar
  x)}~~~~~~~~~~~~~~~~~~~~~~~~~~~~~~
~~~~~~\\ \nonumber
~~~~~~~~~~~~~~~~~~~~~~-\frac{l\,e^{i\pi\theta}+r}{\sqrt{\Lambda}}\,
e^{-\phi(x)}\,e^{-\phi^*(\bar x)}-
\frac{l\,e^{-i\pi\theta}+r}{\sqrt{\Lambda}}\,
e^{-\phi^*(x)}\,e^{-\phi(\bar x)}~,
\end{eqnarray}
where $\Lambda=l^2+r^2+2lr\cos\pi\theta-\sin^2\pi\theta$ and 
\begin{equation}\label{e^-phi(x)=}
e^{-\phi(x)}=\left(\frac{m\,\sqrt{\Lambda} }
{\xi'(x)\,\theta\,\sin\pi\theta}\right)^\frac{1}{2}\,
\,e^{-\frac{i\theta\,\xi(x)}{2}-\frac{i\pi}{2}}~.
\end{equation}
Here $\phi(x)$ is the same field  as it stands in (\ref{ff}), but for imaginary
$p$. This field is not real: $\phi(x)=\phi_1(x)+i\phi_2(x)$, 
and  its real and imaginary parts are related by
\begin{equation}\label{2phi_2'=e^2phi_1}
\phi_2'(x)=\frac{m\,\sqrt{\Lambda}}{2\sin\pi\theta}\,\,e^{2\phi_1(x)}~.
\end{equation} 
It has to be mentioned that the time evolution for imaginary $p$ has 
an oscillating character and there are no asymptotic fields any more.
But the continuation of $in$ and $out$ fields
to the sector of bound states exist and they  become complex conjugated
to each other.

Quantization of this situation is a non-trivial problem indeed. 
At least we do not know global canonical variables, or
an analog of the infinite dimensional translation symmetry to apply an 
alternative quantization scheme (like 
geometric, coherent state, $etc$). It is a typical situation  for
the elliptic monodromy.
What remains still valid from the scattering picture (hyperbolic
monodromy) is the Fourier mode
expansion of  $\phi(x)$  (\ref{phi-modes}) and the Poisson
brackets (\ref{PB,a_n}). But these brackets are not canonical,
since the conjugation rule $a_n^*= a_{-n}$ is violated.
It also remains the parameterization (\ref{V-phi}) and the formula 
for the screening charge (\ref{A}) (with $p=i\theta$). 

Quantization of the zero mode sector can be done by the scheme
described above. The problem is in the non-zero mode sector. Here
one can apply a formal 
algebraic quantization based on the commutation relations (\ref{ccr})
between the $a_n$ operators.
Choosing the normalized vacuum states by
\begin{equation}\label{thata-vacuum}
a_n|\theta_m,\,0\rangle=0~,~~~~~~~~~~  n>0~,
\end{equation} 
where $\theta_m$ are the discrete levels,  
the space of states 
can be constructed similarly to the Fock space, by the action of $a_n$,
with  $n<0$ on $|\theta_m,\,0\rangle$. 
Using the same ordering as in the hyperbolic case, 
formally one gets  the desired structure (locality, conformal weight, $etc$)
for the operator $V$. 
What remains to be specified to calculate matrix elements 
is the scalar product.
Taking into account the above mentioned conjugation
between the `$in$'-`$out$' fields, one gets $a_{-n}^+=b_n$,
where $b_n$ are the Fourier modes for the `$out$-field'  operator.
Thus, the scalar product in the sector of bound states is
given by the continuation (in $p$) of the corresponding matrix elements
of the operator $S_p$. It means that the construction of the $S$-matrix
in a closed form could be the key point for a complete 
understanding of quantization of the elliptic sector.

\section{Conclusions}
We have completed the construction of the basic operator $V=e^{-\varphi}$ in
terms of a free chiral field, started in our previous paper \cite{DJ}. By
going the first steps in the construction of generic
$V_{\alpha}=e^{2\alpha\varphi}$, we were able to relate the mass and boundary
parameters $m_b,l_b,r_b$ in our quantum version of $V$ to those $(m,l,r)$
appearing in the equation of motion and the boundary conditions for the
Liouville field on operator level. Both equations keep the same form as in the
classical case. For the boundary condition the necessary Liouville exponential
has to be understood as a limit of the corresponding bulk operator after
dividing out a short distance singularity.

Some new techniques for the extraction of the spectrum and the definition
of expectation values in states which make sense 
only after some regularization 
and continuation have been developed and tested for a particle in the
Morse  potential. Here complete and exact results are available from the 
solutions of the corresponding  Schr\"odinger equation.
Due to the similarity of the
potential to that in BLT, this particle model should share some qualitative
properties with BLT.  In particular, depending on the sign 
of the parameter $\lambda$,  the Morse particle can have bound and scattering
states or scattering states only. 

The analog of $V$ in the scattering sector has been expressed in terms of a
free canonical $(p,q)$ pair and then continued to the bound state sector.
Then, by requiring Hermiticity and positivity for $V$, as well as an energy 
spectrum bounded from below, the ground state energy can be read off
from the zero of a certain function appearing in this free field
representation of $V$. Higher levels are obtained by the action of the raising
operator $U_+$, which is a part of $V$.  
Identifying parts of $V$ as ingoing and outgoing
exponentials of the particle position, a functional relation for the reflection
amplitude has been derived and solved by turning it into an infinite order
differential equation. The spectrum is then reproduced from the zeros of the
reflection amplitude.

Furthermore, to mimic the problems encountered with the $SL(2,\rr)$ invariant
state in BLT, we have shown that the norm of the state, obtained by continuing
the $\delta$-function normalized scattering states to that imaginary momentum
corresponding to the ground state energy, can be obtained also via
a regularization and continuation process  using either insertions of generic
$V_{\alpha}$ or regularizations of the scattering states. 

Turning back to BLT our interest concentrated on the issue of the spectrum
of highest weight states for the Verma modules (spectrum of the zero mode
contribution to the energy-momentum tensor)
constituting BLT and on the
calculation  of correlation functions. For both issues the analysis
is determined by the projection $\hat V$ of the operator $V$ to the vacuum 
sector with respect to the free field oscillator Fock space.

This projection has been expressed in terms of exponentials of the zero mode
$(p,q)$ pair and two functions of $F(p)$ and $G(p,\sigma)$. Besides the
$\sigma$-dependence and the more complicated form of $F$ and $G$ this
resembles to what has been done for the Morse particle. 

Then from the analysis of $F(p)$ and $G(p,\sigma)$ we determined the wanted
spectrum first for $b^2\ll 1$ and more general for $b^2<1/3$. For values of the 
boundary parameters $\beta _l,~\beta _r <1-b^2$ only one series $\theta _n$
with constant spacing $\hbar =2b^2$ contributes, in agreement with \cite{T}.
However, for values above the critical $1-b^2$ three (for $l=r$) or even four
(for $l\neq r$) such series may contribute. 

The critical case $l_b=r_b=-\cos\pi b^2$ turned out of particular
interest. There the $SL(2,\rr)$ invariant vacuum state at $\theta _0=-bQ$ is
invariant under the action of $\hat V$, and one has the option to restrict the
theory to the corresponding Verma module (ZZ case \cite{ZZ}) or to include it
as a ground state in the theory, which otherwise is nothing else than generic
BLT (FZZT case) just at this special value of the boundary parameters.  

The reflection amplitude for BLT has been derived from the knowledge of $\hat
V$ applying the technique demonstrated for the Morse particle. Thereby an
integral representation for the phase of the reflection was obtained,
which results in a representation in terms of the Barnes Double Gamma
function. The result agrees with the boundary two point function obtained
in the bootstrap approach \cite{FZZ}. In addition, it has been shown that
making use of the freedom to add certain periodic factors, not fixed by
the defining functional relation, modified reflection amplitudes can be
constructed, whose pattern of zeros completely reproduces the spectrum
obtained via the properties of $\hat V$ before. 

A last issue concerns the 1-point function for $V$ with respect to the
$SL(2,\rr)$ invariant vacuum. In the critical case  $l_b=r_b=-\cos\pi b^2$
this quantity is a priori well defined, since the state is in the spectrum
and therefore normalizable. There we get just the ZZ 1-point function 
\cite{ZZ}. In the generic case some regularization and continuation
has to be implemented, which requires further analysis.
\vspace{3mm}

The main part of our results has been presented by G.J. at the conference
``Conformal Field Theory and Integrability: from Condensed Matter Physics
to String Theory'' Yerevan and Tbilisi, 01.10.-09.10. 2007.

\vspace{6mm}

\noindent {\bf {\large Acknowledgments}}

\noindent 
G.J. would like to thank the brothers Zamolodchikov for helpful discussion.\\ 
He thanks ICTP Trieste for hospitality, where a part of
his work was done, RFBR and also GNSF for the grant ST06/4-050.\\
The research was supported in part by DFG with the grant DO 447-4/1.


\newpage

.

\setcounter{equation}{0}

\def\theequation{A.\arabic{equation}}

\noindent{\bf \large{Appendix A}}

\vspace{3mm}

\noindent{\bf Calculation of the coefficients  $c^{l,k}_p(\alpha)$}

\vspace{3mm}

\noindent
In this Appendix we present two different schemes of calculation
of $p$-dependent coefficients $c^{l,k}_p(\alpha)$,
for the expansion (\ref{V_alpha1}).
These calculations are used in Section 2
to find the relation between the parameters ($m_b$, $l_b$, $r_b$) and 
$(m,~l,~r)$. 

First we consider the scheme based on eq. (\ref{VV=VV}).
By (\ref{V_alpha}), this equation reads
\begin{eqnarray}\label{VV=VV1}
\Psi_\alpha(x,-x){\cal V}_\alpha(x,-x)\,
\Psi_{-\frac{1}{2}}(y,-y){\cal V}_{-\frac{1}{2}}(y,-y)=\\ \nonumber
\Psi_{-\frac{1}{2}}(y,-y){\cal V}_{-\frac{1}{2}}(y,-y)
\,\Psi_\alpha(x,-x){\cal V}_\alpha(x,-x)~.
\end{eqnarray}
Using the exchange relations (\ref{exch-p})-(\ref{A-A^}) and the locality
for free field exponentials 
\begin{equation}\label{PsiPsi=PsiPsi}
\Psi_\alpha(x,-x)\,
\Psi_{-\frac{1}{2}}(y,-y)=
\Psi_{-\frac{1}{2}}(y,-y)\,\Psi_\alpha(x,-x)~,
\end{equation}
one can cancel the free-field exponents in (\ref{VV=VV1}) and  reduce
the equation to
\begin{equation}\label{VV=VV2}
\tilde{\cal V}_\alpha(x,-x;y,-y)\,
{\cal V}_{-\frac{1}{2}}(y,-y)=
\tilde{\cal V}_{-\frac{1}{2}}(y,-y;x,-x)
\,{\cal V}_\alpha(x,-x)~.
\end{equation}
Here $\tilde{\cal V}_\alpha(x,-x;y,-y)$ denotes the operator which comes
from ${\cal V}_\alpha(x,-x)$ after its exchange  with 
$\Psi_{-\frac{1}{2}}(y,-y)$. This procedure shifts the index $p$ 
of the coefficients $c^{l,k}_p(\alpha)$ by $p\mapsto p+2ib^2$ and
creates new screening charges $A(y)$ and $A(-y)$
according to (\ref{exch-2}). Similarly, the
exchange of ${\cal V}_{-\frac{1}{2}}(y,-y)$ with $\Psi_\alpha(x,-x)$
creates $\tilde{\cal V}_{-\frac{1}{2}}(y,-y;x,-x)$.
Thus, both sides of eq. (\ref{VV=VV2}) are power series in
$A(x)$, $A(-x)$, $A(y)$ and $A(-y)$. Comparing the coefficients of 
the corresponding monoms we get equations for the coefficients 
$c^{l,k}_p(\alpha)$.
For example, the terms linear in the screening charges provide four equations.
One of them, which comes from the coefficients of $A(x)$, is
\begin{equation}\label{f=f}
    e^{2i\pi b^2}\,c^{1,1}_{p+i\hbar}(\alpha)=c^{1,1}_{p}(\alpha)
+\mu^-_p(\alpha)
\left(c^{1,0}_{p-4i\alpha b^2}+c^{1,1}_{p-4i\alpha b^2}\right)~.
\end{equation}
Here $c^{1,0}_{p}$ and $c^{1,1}_{p}$ are given by (\ref{f-g,M-N}) and
$\mu^-_p(\alpha)$ is the coefficient of $A(y)$ in (\ref{exch-2})
for $\epsilon(x-y)=-1:$
\begin{equation}\label{lambda_p}
\mu^-_p(\alpha)=\frac{i\sin(2\pi\alpha b^2)\,\,e^{-\pi
(p-2i\alpha b^2-ib^2)}}
{\sinh\pi(p-2i\alpha b^2-ib^2)}~.
\end{equation}
Eq. (\ref{f=f}) is solved by (\ref{f^alpha_p}). The comparison
of the coefficients of $A(-x)$ leads to a similar equation for 
$c^{1,0}_{p}(\alpha)$, which is solved by (\ref{g^alpha_p}).
Then one checks that the two other equations,
provided by the coefficients of $A(y)$ and $A(-y)$ are also solved
by (\ref{g^alpha_p})-(\ref{f^alpha_p}).

Considering the coefficients of $A(-x)A(x)$,
one finds the equation for $c^{2,1}_{p}(\alpha)$:
\begin{equation}\label{N=N}
c^{2,1}_{p+i\hbar}(\alpha)=c^{2,1}_{p}(\alpha)
+\nu_p(\alpha)~,
\end{equation}
Here $\nu_p(\alpha)$ is calculated by the  
coefficients $c^{1,k}_{p}$ $(k=0,1)$, $c^{2,k}_{p}$ $(k=1,2)$
and $c^{1,k}_{p}(\alpha)$ $(k=0,1)$, which are already known from
(\ref{f-g,M-N})-(\ref{M-N}) and (\ref{g^alpha_p})-(\ref{f^alpha_p}).
The calculation of $\nu_p(\alpha)$ is a rather long, but straightforward
procedure. Finally, eq. (\ref{N=N}) is solved by (\ref{N^alpha_p}).

Note that the solutions of the equations
(\ref{f=f}) and (\ref{N=N}), have a freedom
in adding the terms $e^{\frac{\pi np}{b^2}}$, with integer $n$.
We neglect this freedom, requiring `smooth' dependence on $\hbar =b^2$.
This question is discussed in Section 3 in more detail, 
where a method of solution of this type of equations is given as well.
\vspace{2mm}

Now we present an alternative scheme based on the construction
of the vertex operators $V_{-\frac{n}{2}}(x,\bar x)$.
These operators are associated with the exponentials
$e^{-n\varphi(x,\bar x)}$. The operator
$V_{-\frac{1}{2}}(x,\bar x)\equiv V(x,\bar x)$ is given by 
(\ref{^V1}) and others can be obtained 
step by step as its regularized powers  
\begin{equation}\label{V^n}
V_{-\frac{n+1}{2}}(x,\bar x)=
\lim_{\epsilon\rightarrow 0} \,
V(x+\epsilon,\bar x-\epsilon)\,
V_{-\frac{n}{2}}(x,\bar x)\,\,|\epsilon|^{n b^2}~.
\end{equation}
The screening charge operator $A(x)$ 
does not create short distance
singularities in the operator products like (\ref{V^n}), and 
the primary free-field 
exponentials (\ref{Psi_alpha}) provide 
\begin{equation}\label{short-dist}
\lim_{\epsilon\rightarrow 0} 
\Psi_\alpha(x,\bar x)\,\Psi_\beta(x+\epsilon,\bar x-\epsilon)
\,|\epsilon|^{{4\alpha
\beta b^2}}=\Psi_{\alpha+\beta}(x,\bar x)~.
\end{equation}
This formula helps to calculate the regularized powers (\ref{V^n})
by recurrence relations.

The operator $V_{-\frac{n}{2}}$ is a polynomial in $m_b$ of the order $2n$.
Assuming that it can be written similarly to
(\ref{^V1})
\begin{eqnarray}\label{^Vn}
V_{-\frac{n}{2}}(x,\bar x)=e^{-\frac{i\pi\,n^2\,b^2}{4}}\,e^{-n\phi(\bar
x)}\,e^{-n\phi(x)}\Big[1+m_b\left(A(\bar x)c^{1,0}_{n,\,p}+A(x)
c^{1,1}_{n,\,p}
\right)+~~~~~~~~~~~~~~~~~~~
\\ \nonumber
m_b^2\left(A^2(\bar x)c^{2,0}_{n,\,p}+A(\bar x)A(x)c^{2,1}_{n,\,p}
+A^2(x)c^{2,2}_{n,\,p}\right)+...\Big]~,~~
\end{eqnarray}
where, for convenience, we denote the expansion coefficients
by $c^{l,k}_{n,\,p}$, instead of $c^{l,k}_p(-\frac{n}{2})$.
Applying then the exchange relations 
(\ref{exch-p})-(\ref{A-A^}) to (\ref{V^n}),
the operator  $~V_{-\frac{n+1}{2}}~$ is reduced to the form
(\ref{^Vn}), and one finds recurrence relations 
for the coefficients
$c^{l,k}_{n,\,p}$. 

For example, the coefficients of $A(\bar x)$ and $A(x)$ 
provide
\begin{equation}\label{g(n+1)}
c^{1,0}_{n+1,\,p}= c^{1,0}_{n,\,p}+e^{-i\pi n b^2}\,
c^{1,0}_{p+2inb^2}+\mu^+_p(-\,n/2)\,c^{1,1}_{p+2in b^2}~,
\end{equation}
and
\begin{equation}\label{f(n+1)}
c^{1,1}_{n+1,\,p}= c^{1,1}_{n,\,p}+e^{i\pi n b^2}\,
c^{1,1}_{p+2inb^2}+\mu^-_p(-\,n/2)\,c^{1,0}_{p+2inb^2}~,
\end{equation}
respectively. Here $c^{1,0}_{p}$ and $c^{1,1}_{p}$ 
are given by (\ref{f-g,M-N}),  
$\mu^-_p(-\,n/2)$ is the coefficient (\ref{lambda_p}) for 
$\alpha=-\,n/2$ and $\mu^+_p(-\,n/2)$ differs from it only
by the sign of the exponent in (\ref{lambda_p}).

These  recurrence relations are solved by
\begin{equation}\label{g(n)}
c^{1,0}_{n,\,p}=\frac{\sin\pi n b^2}{\sin\pi b^2}\,\,
\frac{l_b\,e^{\pi p}\,e^{i\pi(n-1)b^2}
+r_b\,e^{-i\pi n b^2}}
{\sinh\pi(p+2inb^2-ib^2)}~,
\end{equation}
\begin{equation}\label{f(n)}
c^{1,1}_{n,\,p}=\frac{\sin\pi n b^2}{\sin\pi b^2}\,\,
\frac{l_b\,e^{-\pi p}\,e^{-i\pi(n-1)b^2}
+r_b\,e^{i\pi n b^2}}
{\sinh\pi(p+2inb^2-ib^2)}~.
\end{equation}
The continuation of these expressions to real $n$ and denoting 
$n= -2\alpha$ leads to (\ref{g^alpha_p})-(\ref{f^alpha_p}).

The same procedure for the coefficient $c^{2,1}_{n,\,p}$
yields a similar recurrence relation 
\begin{equation}\label{c^21}
c^{2,1}_{n+1,\,p}= c^{2,1}_{n,\,p}+X_{n,\,p}
\end{equation}
where $X_{n,\,p}$ is defined by the already known 
coefficient. Its calculation
is similar to the calculation of $\nu_p(\alpha)$ in (\ref{N=N}).
As a result, the solution of (\ref{c^21})
coincides with (\ref{N^alpha_p}) for $\alpha=-\,n/2$.

The higher order coefficients of the expansion (\ref{V_alpha1}) can also
be calculated by the schemes 
presented here, but a closed form of the general
coefficient $c^{l,k}_{n,\,p}$ is still missing.

 \vspace{6mm}

\setcounter{equation}{0}

\def\theequation{B.\arabic{equation}}

\noindent{\bf \large{Appendix B}}

\vspace{3mm}

\noindent{\bf Particle in the Morse Potential}

\vspace{3mm}

\noindent {\bf 1. The eigenstates and the reflection amplitude}

\vspace{3mm}

\noindent The  Schr\"odinger equation in the Morse potential
(\ref{Morse})
\begin{equation}\label{Schr}
-\frac{\hbar^2}{2}\Psi_E''(y)+\left[2m^2\,e^{2y}+2m\lambda\,
e^y\right]\Psi_E(y)=E\Psi_E(y)~
\end{equation}
can be reduced \cite{FL} to the equation for the confluent
hypergeometric functions and the eigenstates for positive
energies ($E>0$) read
\begin{eqnarray}\label{Psi}
\Psi_E(y)&=&\left(\frac{4m}{\hbar}\right)^{ik}\,
\frac{\Gamma(-2ik)}{\Gamma\left(
 \nu-ik\right)}\,e^{-\frac{z}{2}}\,\Phi\left(
 \nu+ik,\,1+2ik,\,z\right)\,e^{iky}+\\ \nonumber
&&\left(\frac{4m}{\hbar}\right)^{-ik}\,
\frac{\Gamma(2ik)}{\Gamma\left(
 \nu +ik\right)}\,\,e^{-\frac{z}{2}}\,\Phi\left(\nu -ik,\,
1-2ik,\,z\right)\,e^{-iky} =
W_{-\frac{\lambda}{\hbar}, \,ik}(z)\, z^{-\frac{1}{2}}~.
\end{eqnarray}
Here $\Phi$ is the confluent hypergeometric function \cite{GR}
\begin{equation}\label{Phi(z)}
\Phi(\alpha,\gamma,z)=1+\frac{\alpha}{\gamma}\,\frac{z}{1!}
+\frac{\alpha(\alpha+1)}{\gamma(\gamma+1)}\,\frac{z^2}{2!}+...~,
\end{equation}
$W_{-\frac{\lambda}{\hbar}, \,ik}(z)$ is the Whittaker's function \cite{GR}  
and we have used the notations
\begin{equation}\label{x,k,mu}
 z=\frac{4m}{\hbar}\,\ e^y,~,~~~~k=\frac{\sqrt{2
E}}{\hbar}~, ~~~~     \nu   =\frac{1}{2}+\frac{\lambda}{\hbar}~.
\end{equation}
The wave functions (\ref{Psi}) vanish at $y\rightarrow\infty$ and
their behavior at $y\rightarrow -\infty$ can be written as
\begin{equation}\label{y=-infty}
\Psi_E(y)\sim a^*(k)e^{iky}+a(k)e^{-iky}~,
\end{equation}
with
\begin{equation}\label{a(k)}
 a(k)=\left(\frac{4m}{\hbar}\right)^{-ik}\,\,
 \frac{\,\Gamma(2ik)}{\Gamma\left(    \nu   +ik\right)}~.
\end{equation}
The ratio of $out$-going and $in$-coming coefficients defines the
reflection amplitude
\begin{equation}\label{Rk}
r_\lambda(k)=\frac{a(k)}{a^*(k)}=
\left(\frac{4m}{\hbar}\right)^{-2ik}\,\, \frac{\Gamma(\,2ik\,)
\,~\Gamma(    \nu   -ik)} {\Gamma(-2ik)\,\Gamma(    \nu   +ik)}~.
\end{equation}
If $    \nu   $ is negative there are, in addition, the bound states at
$k=i\kappa_n$, where $\kappa_n=    \nu   +n$. The second term in
(\ref{Psi}) vanishes for these imaginary values of $k$ and,
therefore, the normalized wave functions can be written as
\begin{eqnarray}\label{psi-n}
\psi_n=c_n\,\,e^{-(    \nu   +n)y}\,\,e^{-\frac{z}{2}}\,\left(1+
\sum_{m=1}^n \, \frac{n(n-1)\cdot\cdot\cdot(n-m+1)}{(2    \nu   +2n-1)
\cdot\cdot\cdot(2    \nu   +2n-m)}\, \frac{z^{m}}{m!}\right)~,
\end{eqnarray}
where $c_n$ are normalization coefficients. The number of the
bound states is restricted by $0\leq n<-    \nu   $. The coefficients $c_n$
can be expressed through the derivatives of $r_\lambda(k)$ at
$k=i\kappa_n$, where this function has simple zeros.     
\cite{N}.
   
For this purpose let us introduce the function
$\Psi_k(y)=\frac{\Psi_E(y)}{a^*(k)}$, which has the following
asymptotic behavior at $y\rightarrow -\infty$
\begin{equation}\label{y=-infty1}
\Psi_k(y)\sim e^{iky}+r_\lambda(k)e^{-iky}~.
\end{equation}
Its analytical continuation is related to the normalized wave
functions (\ref{psi-n}) by
\begin{eqnarray}\label{psi-n=}
\psi_n=c_n\,\Psi_{i\kappa_n}~.
\end{eqnarray}
From the Schr\"odinger equation (\ref{Psi})
one finds that
\begin{equation}\label{Psi-dotPsi}
{\Psi_k}''(y)\,\partial_k\Psi_k(y)-
\partial_k\Psi_k''(y)\Psi_k(y)=2k\Psi_k^2(y)~,
\end{equation}
and its integration leads to
\begin{equation}\label{int-Psi^2}
2k\,\int_{y}^{\infty}\Psi_k^2(y_{_1})\,dy_{_1}=
\partial_k\Psi_k'(y)\Psi_k(y)
-{\Psi_k}'(y)\,\partial_k\Psi_k(y)~.
\end{equation}
Taking here the limit $y\rightarrow -\infty$ at $k=i\kappa_n$, we obtain
\begin{equation}\label{Psi-Psi}
\langle\Psi_{i\kappa_n}|\Psi_{i\kappa_n}\rangle=-i\partial_k
r_\lambda(k)|_{k=i\kappa_n}~,
\end{equation}
which by (\ref{Rk}) provides
\begin{equation}\label{cn}
c_n=\left(\frac{\hbar}{4m}\right)^{    \nu   +n}\,\,
\sqrt{\frac{(-)^n\,\Gamma(2    \nu   +2n)}{n!\,
\Gamma(-2    \nu -2n)\,\Gamma(2    \nu   +n)}}~.
\end{equation}
Finally, we present the normalized wave functions $\psi_0$ and $\psi_1$, 
written in terms of $z$ (see (\ref{x,k,mu}))
\begin{equation}\label{psi0-1}
\psi_{_0}=\frac{1}{\sqrt{\Gamma(2|    \nu   |)}}\,\,
z^{|    \nu   |}\,e^{-\frac{z}{2}}~, ~~~~~ \psi_{_1}=
\sqrt{\frac{2|    \nu   |-1}{\Gamma(2|    \nu   |-2)}}\,\,\,
z^{|    \nu   |-1}\,e^{-\frac{z}{2}}\left(1-\frac{z}{2|    \nu   |-1}\right)~.
\end{equation}

 \vspace{5mm}

\noindent{\bf 2. The dynamical equation for the operator $V(\tau)$}

\vspace{3mm}

\noindent The Hamiltonian
$H=\frac{1}{2}\,p_y^2+2m^2\,e^{2y}+2m\lambda\, e^y$ provides the
following time evolution equations for the Heisenberg operators
$V=e^{-y}$ and $p_y$
\begin{equation}\label{dotV}
\dot V= -V\left(p_y+\frac{i\hbar}{2}\right)~, ~~~~~~~~~ \dot
p_y=-\left(\frac{4m^2}{V^2}+\frac{2m\lambda}{V}\right)~.
\end{equation}
The second derivative of $V$ then becomes
\begin{equation}\label{ddotV}
\ddot V=
V\left(p_y+\frac{i\hbar}{2}\right)^2+\frac{4m^2}{V}+2m\lambda~,
\end{equation}
and expressing the momentum operator $p_y$ from (\ref{dotV}) through
$V$ and $\dot V$ one finds
\begin{equation}\label{ddotV1}
\ddot V= \dot V\,\frac{1}{V}\,\dot V+\frac{4m^2}{V}+2m\lambda ~.
\end{equation}
From (\ref{dotV}) also follows the commutator $\dot V V-V\dot
V=i\hbar\, V^2$,
which together with (\ref{ddotV1}) lead to the dynamical equation
(\ref{V-qu-eq}).

 \vspace{6mm}

 \setcounter{equation}{0}

\def\theequation{C.\arabic{equation}}

\noindent{\bf \large{Appendix C}}

\vspace{3mm}

\noindent {\bf Integral representation of $\gamma(p)$}

\vspace{3mm}

\noindent
In this appendix we calculate the phase  $\,\gamma(p)=i\hbar\,\log R(p)$
of the reflection amplitude of BLT. It 
satisfies the equation 
\begin{equation}\label{gamma'}
\gamma^{\,\prime}(p)=\hat O_\hbar\,\log F(p)~,
\end{equation}
where $\hat O_\hbar$
is the operator (\ref{^O1}) and $\log F(p)$ is obtained from 
(\ref{F(p)=}), (\ref{up=}). 
We follow the scheme described in Section 2, using  $2b^2$
instead of $\hbar$.

The function $\log F(p)$ can be written as
\begin{equation}\label{log-F}
\log F(p)=\log\left(4\pi^2\,m_b^2\,\Gamma^2(1+b^2)\right)-
\log(p^2+b^4)+\log F_1(p)-\log F_2(p)~,
\end{equation}
with 
\begin{equation}\label{F1}
F_1(p)=\Gamma(ip)\,\Gamma(-ip)
\,\Gamma(1-b^2+ip)\,\Gamma(1-b^2-ip)~~~
\end{equation}
and
\begin{equation}\label{F2}
F_2(p)={\prod_{\varepsilon=\pm 1,\,\nu=\pm}
\Gamma\left(\frac{1}{2}-\frac{\varepsilon\,\beta_\nu}{2}+
\frac{ip}{2}\right)
\Gamma\left(\frac{1}{2}-\frac{\varepsilon\,\beta_\nu}{2}-
\frac{ip}{2}\right)}~.
\end{equation}
Note that $F_2(p)$ corresponds to the product of four $\cos$-terms 
in (\ref{up=}), expressed through the $\Gamma$-function by
(\ref{Gamma-Gamma=ch}). 
The integral representations (\ref{log=})
and (\ref{log-Gamma=}) then lead to
\begin{equation}\label{logF=}
\log F(p)=\log\left(4\pi^2\,m_b^2\,\Gamma^2(1+b^2)\right)+
\int_0^\infty
dt\,\left[A(t)\,e^{ipt}+B(t)\,e^{-ipt}+C(t)\right]~.
\end{equation}
The function $A(t)$ is represented as a sum $A(t)=A_0(t)+A_1(t)+A_2(t)$,
where $A_0(t)$, $A_1(t)$ and $A_2(t)$ correspond to the contributions  
from $-\log(p^2+b^4)$, $\log F_1(p)$ and $-\log F_2(p)$, 
respectively.
Writing $A_2(t)$ also as a sum 
$A_2(t)=\sum_{\varepsilon,\,\nu} A_{2}^{\varepsilon,\,\nu}(t)$ 
corresponding to (\ref{F2}), we
read off these functions by (\ref{log=}) and (\ref{log-Gamma=}).
We do the same for $B(t)$ and $C(t)$ and obtain  
\begin{equation}\label{A0(t)=}
A_0(t)=\frac{e^{-b^2\,t}}{t}=B_0(t)~,~~~~~~~~~~~~~~~~
~~~~~~~~
C_0=-\frac{2e^{-t}}{t}~,~~~~~~~~~~~~~~~~~~~~~~~~~~~~~~~
\end{equation}
\begin{equation}\label{A1(t)=}
A_1(t)=\frac{1+e^{-(1-b^{2})t}}{t\left(1-e^{-t}\right)}=B_1(t)~,
~~~~~~~~~~~~~~
C_1=-\frac{2e^{-t}}{t}\left(1+b^{2}+\frac{2}{1-e^{-t}}\right)
~,~~~~~
\end{equation}
\begin{equation}\label{A2(t)=}
A_{2}^{\varepsilon,\,\nu}(t)=
-\frac{e^{-(1-\varepsilon\,\beta_\nu)t}}
{t\left(1-e^{-2t}\right)}=B_{2}^{\varepsilon,\,\nu}(t)~,
~~~~~~~~
C_{2}^{\,\varepsilon,\,\nu}=
\frac{2e^{-2t}}{t\left(1-e^{-2t}\right)}+(1+\varepsilon\,\beta_\nu)\,
\frac{e^{-2t}}{t}~.~
\end{equation}
Then, similarly to (\ref{gamma=}), we find a solution of (\ref{gamma'})
in the form
\begin{equation}\label{gamma(p)=}
\gamma(p)=2\pi b^2+p\log\left(4\pi^2\,m_b^2\,
\Gamma^2(1+b^2)\right)+\int_0^\infty dt\,\left[\tilde
A(t)\,e^{ipt}+\tilde B(t)\,e^{-ipt}+p C(t)\right]~,
\end{equation}
where $\tilde A(t)$ and $\tilde B(t)$ are related to $A(t)$
and $B(t)$ of (\ref{logF=}) by (\ref{A(t),B(t)->}). The integral term in
(\ref{gamma(p)=}) corresponding to $\log F_1(p)-\log(p^2+b^4)$ contains the
integral $I_3$ of (\ref{I-2}), and it cancels the constant term $2\pi b^2$. 
Finally (\ref{gamma(p)=})
leads to (\ref{gamma(p)=1}).

The reflection amplitude $R(p)=e^{-\frac{i}{\hbar}\gamma(p)}$ corresponding to
(\ref{gamma(p)=1}) 
can be expressed through the Barnes Double Gamma function \cite{Ba}  
$\Gamma_b(z)$ (see (\ref{R(p)=})), 
since for $\mbox{Re}\,\, z>0$ the following integral representation holds
\cite{FZZ}
\begin{equation}\label{Gamma-b}
\log\Gamma_b(z)=\int_0^\infty \frac{dt}{t}\,\left[\frac{e^{-zt}-
e^{-\frac{1}{2}\left(b+\frac{1}{b}\right)t}}
{\left(1-e^{-bt}\right)\left(1-e^{-\frac{t}{b}}\right)}
-\frac{\left(b+\frac{1}{b}-2z\right)^2}{8}\,e^{-t}-
\frac{b+\frac{1}{b}-2z}{2t}\right]~.
\end{equation}
The continuation of this integral to  $\mbox{Re}\,z\leq 0$ 
can be done by the
relations 
\begin{equation}\label{Gamma_b(z+1/b)}
\Gamma_b(z+b)=\sqrt{\frac{2\pi}{ b}}\,\,\frac{b^{bz}}{\Gamma(bz)}\,\,
\Gamma_b(z)~.
~~~~~~~~~
\Gamma_b(z+1/b)=\frac{\sqrt{2\pi b}}{\Gamma(z/b)\,b^{\frac{z}{b}}}\,\,
\Gamma_b(z)~,
\end{equation}
which easily follow from (\ref{Gamma-b}).

\vspace{6mm}

 \setcounter{equation}{0}

\def\theequation{D.\arabic{equation}}

\noindent{\bf \large{Appendix D}}

\vspace{3mm}

\noindent {\bf A list of useful formulas from \cite{GR} used in the paper}


\begin{equation}\label{Gamma-Gamma=ch}
\Gamma\left(\frac{1}{2}+\theta\right)\Gamma\left(\frac{1}{2}-\theta\right)=
\frac{\pi}{\cos\pi\theta}~,~~~~~~
\Gamma\left(1+ip\right)\Gamma\left(1-ip\right)=
\frac{\pi p}{\sinh\pi p}~.
\end{equation}
\begin{eqnarray}\label{F(1/z)}
_2F_1(a,b,c;z)=\frac{\Gamma(b-a)\Gamma(c)}{\Gamma(b)\Gamma(c-a)}\,
\left(-{z}\right)^{-a}\,_2F_1(a,1-c+a,1-b+a;1/z)
~~~~~~~~~~~~~~~~~
\\ \nonumber ~~~~~~~~+
\frac{\Gamma(a-b)\Gamma(c)}{\Gamma(a)\Gamma(c-b)}\,
\left(-{z}\right)^{-b}\,_2F_1(1-c+b,b,1-a+b;1/z)~
~~~(|\mbox{arg}\,(-z)|< \pi)~.~~ 
\end{eqnarray}
\begin{eqnarray}\label{F(1-z)}
_2F_1(a,b,c;z)=\frac{\Gamma(c-a-b)\Gamma(c)}{\Gamma(c-a)\Gamma(c-b)}\,\,
_2F_1(a,b,a+b-c+1;1-z)
~~~~~~~~~~~~~~~~~~~~~~~~~
\\ \nonumber ~~~~~~+
\frac{\Gamma(a+b-c)\Gamma(c)}{\Gamma(a)\Gamma(b)}\,
\left(1-z\right)^{c-a-b}\,_2F_1(c-a,c-b,c-a-b+1;1-z)~.
\end{eqnarray}
\begin{equation}\label{log=}
\log\left(p^2+a^2\right)=-\int_0^\infty
\frac{dt}{t}\left[\left(e^{ipt}+e^{-ipt}\right)e^{-at}
-2e^{-t}\right]~~~~~(p>0,~a\geq 0).
\end{equation}
\begin{equation}\label{log(a)}
~~~~~~~~~~\log a=\int_0^\infty \frac{dt}{t}\left(e^{-t}-e^{-at}\right)~
~~~~~~~~~~~~~~~~~~~~~~~~~~~~~~~~~~(a>0).
\end{equation}
\begin{equation}\label{log-Gamma=}
~~~~~~~\log \Gamma(z)=\int_0^\infty \frac{dt}{t}\left[
\frac{e^{-zt}-e^{-t}}{1-e^{-t}} +(z-1)e^{-t}\right]~~~~~~~~~~~
(\mbox{Re}\,z> 0)~.
\end{equation}
\begin{equation}\label{int}
~~~~~~~~~~\int_0^{2\pi}dy\,e^{\,\rho\,
y}\,\left(1-e^{\pm\,iy}\right)^\alpha=
\frac{2\pi\,\Gamma(1+\alpha)\,e^{\,\pi\,\rho}}
{\Gamma(1\pm i\rho)\Gamma(1+\alpha\mp i\rho)}~~~~~~~~~~~(\alpha >-1)~.~~~~
\end{equation}
\begin{eqnarray}\label{log-Gamma=1}
\log \Gamma(z)=\left(z-\frac{1}{2}\right)\log z-z+\log\sqrt{2\pi}+
\int_0^\infty \frac{dt}{t}\,e^{-zt}\,
\left(\frac{1}{e^t-1}-\frac{1}{t}+\frac{1}{2}\right)\\ \nonumber
(\mbox{Re}\,z\geq 0)~.\\
\label{stirling}
\log\Gamma(z)=\left(z-\frac{1}{2}\right)\log z-z+\log\sqrt{2\pi}+
\sum_{n\geq1}^m\frac{B_{2n}}{2n(2n-1)\,z^{2n-1}}+O(z^{2m+1})\\ \nonumber
(|\mbox{arg}\,z|< \pi)~.
\end{eqnarray}
\begin{equation}\label{log(2pi)}
\int_0^\infty \frac{dt}{t}\left(\frac{2}{t}-
\frac{2e^{-t}}{1-e^{-t}} -e^{-t}\right)=\log(2\pi)~.
\end{equation}
\begin{equation}\label{int(We)}
\int_0^\infty \frac{dz}{z} \,\,e^{-\frac{z}{2}}\,z^\nu\,W_{a,b}(z)=
\frac{\Gamma(\nu+\frac{1}{2}-b)\,\Gamma(\nu+\frac{1}{2}+b)}{\Gamma(\nu+1-a)}~
~~~~~~~~(\mbox{Re}\,(\nu+1/2\pm b)>0)~.
\end{equation}
\begin{equation}\label{Gamma(x+iy)}
\lim_{|y|\rightarrow\infty}|\Gamma(x+iy)| \,e^{\frac{\pi|y|}{2}}\,\,
|y|^{\frac{1}{2}-x}=\sqrt{2\pi}~~~~~~~~(x~~~ \mbox{and}~~~y~~~
\mbox{are real})~.
\end{equation}

\end{document}